\documentclass[11pt]{article}
\pdfoutput=1 % if your are submitting a pdflatex (i.e. if you have
\usepackage{jheppub} % for details on the use of the package, please
\usepackage{braket,slashed,bm,wrapfig}
\usepackage{array,multirow}
\usepackage[normalem]{ulem}
\usepackage[T1]{fontenc} % if needed
\usepackage{graphicx,subfig}
\usepackage{mathrsfs}

\newcommand{\pref}[1]{{(\ref{#1})}}

\newcommand{\ssB}{{\scriptscriptstyle B}}
\newcommand{\ssH}{{\scriptscriptstyle H}}

\newcommand{\ssL}{{\scriptscriptstyle L}}
\newcommand{\ssS}{{\scriptscriptstyle S}}
\newcommand{\ssQ}{{\scriptscriptstyle Q}}
\newcommand{\ssW}{{\scriptscriptstyle W}}
\newcommand{\ssY}{{\scriptscriptstyle Y}}

\newcommand{\SM}{{\scriptscriptstyle SM}}
\newcommand{\HI}{{\scriptscriptstyle HI}}
\newcommand{\EF}{{\scriptscriptstyle EF}}
\newcommand{\EFT}{{\scriptscriptstyle EFT}}
\newcommand{\cO}{{\cal O}}

\newcommand{\nn}{\nonumber}

\newcommand{\cL}{\mathcal L}
\newcommand{\cH}{\mathcal H}

\def\rd{{\rm d}}

\def\nn{\nonumber\\ }

\begin{document}

\def\bea{\begin{eqnarray}}
\def\eea{\end{eqnarray}}
\def\be{\begin{equation}}
\def\ee{\end{equation}}

\def\endignore{}
\def\ignore #1\endignore{} % use to "comment out" text

\newcommand{\roughly}[1]{\mathrel{\raise.3ex\hbox{$#1$\kern-0.85em
\lower1ex\hbox{$\sim$}}}}
\newcommand{\lsim}{\roughly<}
\newcommand{\gsim}{\roughly>}

\newcount\hour \newcount\minute
\hour=\time \divide \hour by 60
\minute=\time
\count99=\hour \multiply \count99 by -60 \advance \minute by \count99
\newcommand{\mydate}{\ \today \ - \number\hour :00}
\preprint{CERN-PH-TH/2014-XXX}
\title{On the Predictiveness of \\Single-Field Inflationary Models}

\author{
C.P. Burgess${}^{1,2}$, Subodh P. Patil${}^{3}$ and Michael Trott${}^3$\\

${}^1$ Department of Physics $\&$ Astronomy, McMaster University, Hamilton ON, Canada\\
${}^2$ Perimeter Institute for Theoretical Physics, Waterloo ON, Canada\\
${}^3$ Theory Division, Physics Department, CERN, CH-1211 Geneva 23, Switzerland}

\abstract{
We re-examine the predictiveness of single-field inflationary models and discuss how an unknown UV completion can complicate determining inflationary model parameters from observations, even from precision measurements. Besides the usual naturalness issues associated with having a shallow inflationary potential, we describe another issue for inflation, namely, unknown UV physics modifies the running of Standard Model (SM) parameters and thereby introduces uncertainty into the potential inflationary predictions. We illustrate this point using the minimal Higgs Inflationary scenario, which is arguably the most predictive single-field model on the market, because its predictions for $A_\ssS$, $r$ and $n_s$ are made using only one new free parameter beyond those measured in particle physics experiments, and run up to the inflationary regime. We find that this issue can already have observable effects. At the same time, this UV-parameter dependence in the Renormalization Group allows Higgs Inflation to occur (in principle) for a slightly larger range of Higgs masses. We  comment on the origin of the various UV scales that arise at large field values for the SM Higgs, clarifying cut off scale arguments by further developing the formalism of a non-linear realization of $\rm SU_L(2) \times U(1)$ in curved space. We discuss the interesting fact that, outside of Higgs Inflation, the effect of a non-minimal coupling to gravity, even in the SM, results in a non-linear EFT for the Higgs sector. Finally, we briefly comment on post BICEP2 attempts to modify the Higgs Inflation scenario.}
\maketitle
%\section{Introduction}
%%%%%%%%%%%%%%%%%%%%%%%

\section{Introduction:}

Recently, the LHC has discovered a Higgs-like boson \cite{Aad:2012tfa, Chatrchyan:2012ufa}, and Planck \cite{Ade:2013uln} has reported precise measurements of the properties of the Cosmic Microwave Background (CMB).\footnote{While this paper was in press, the even more exciting announcement of a measurement of $r$ was made by the BICEP2 collaboration \cite{Ade:2014xna}, we briefly comment on this development in the context of Higgs Inflation in a note added in the conclusions.} In both cases, simplicity apparently rules. The LHC results are consistent with the Standard Model (SM), including the simplest linear realization of $\rm SU_L(2) \times U_Y(1)$ in the scalar sector, and rule out many exotic alternatives. The properties of the CMB as inferred by Planck, WMAP \cite{Hinshaw:2012aka} and other ground based observations \cite{Hou:2012xq,  Sievers:2013ica} are consistent with the Gaussian, adiabatic primordial curvature perturbations, typically predicted by single-field slow-roll models. This seemingly rules out many more exotic inflationary scenarios\footnote{Although some of the apparently simplest scenarios, such as some power law single field models are also disfavoured by Planck data.}.

Both developments raise the stakes for the Higgs Inflation (HI) proposal \cite{Bezrukov:2007ep}\cite{DeSimone:2008ei,Bezrukov:2008ej} which aspires to use the SM Higgs boson as the single-field inflaton. The idea is to do so by adding the term $\delta \mathcal{L} = -\xi (H^\dagger H) \, R$ to the combined Einstein-Hilbert and SM Lagrangians (where $H$ is the Higgs doublet and $R$ is the metric's Ricci curvature scalar), thereby making the Higgs sector into a non-minimally coupled inflationary model \cite{Spokoiny:1984bd, Salopek:1988qh}. This seems a very benign, and arguably simple modification of known physics, since the new term is proportional to a dimensionless coupling ($\xi$) that is allowed by the symmetries given the  SM field content.

At face value this model has many compelling features, no new fields are required beyond those describing particles now known to exist. Furthermore, it seems extremely predictive because all parameters except $\xi$ are determined by non-cosmological physics, and $\xi \simeq 10^4$ is fixed by requiring the amplitude of primordial scalar fluctuations agree with CMB observations. Once this is arranged, the predictions for the scalar spectral index, $n_s$, and primordial tensor-to-scalar ratio, $r$, become parameter-independent at leading order (other than the dependence on SM parameters arising in the reheating analysis that is used to fix the number of inflationary $e$-foldings, $N_e$). Best yet, the predictions are successful: $n_s \simeq 0.967$ and $r \simeq 0.0031$ agree well with the Planck data.\footnote{With the advent of BICEP2's measurement of $r = 0.20^{+ 0.07}_{-0.05}$ the later prediction is in conflict with the data. But, the Higgs inflation paradigm has since been modified {\it post-hoc} to accommodate a larger $r$. See the comments at the end of the paper regarding this development.} See \cite{Bezrukov:2013fka} for a recent review on this model.

This success, and the improved observational constraints, has led to a more systematic assessment of inflationary models in view of the observations \cite{Martin:2013nzq}, with the HI model used as the benchmark model against which others are assessed in a Bayesian comparison. Indeed, such an analysis favours models for which inflation is not ruined by small parameter changes, and whose $n_s$ and $r$ predictions agree with the data as their parameters vary over a wide range of values. This tends to reward models with exponential potentials, like $V(\phi) = A - B e^{-\lambda \phi}$, for which the slow-roll condition requires only that $\phi$ be sufficiently large. This includes both the HI model and $R^2$ inflation \cite{Starobinsky:1980te}\footnote{See \cite{Kehagias:2013mya} for a study of their essential equivalence in the large field regime.}. This result can also be viewed to be consistent with many models where exponentials arise in higher-dimensional theories, where the inflaton is a geometrical modulus (like the size, $r$, of an extra dimension) given that the associated energies can arise as powers of $1/r$ and the canonical field for such a quantity is $\phi \sim \ln r$ \cite{Burgess:2013sla, Roest:2013fha}. In particular, Ref \cite{Burgess:2001vr}
advocated extra dimensional models for exactly this exponential behavior far in advance of Planck data.

In this paper, we re-examine the predictiveness of single-field inflationary models, using the HI model as their poster child. We revisit the issue of the sensitivity of inflationary predictions to unknown UV physics, with the effects of this physics systematized within an Effective Field Theory (EFT) framework \cite{Donoghue:1995cz, Burgess:2003jk}  for gravity. Beyond the `usual' UV sensitivity issues that are well known: the propensity of UV physics to ruin the flatness of the inflaton potential; and the sensitivity of slow-roll parameters to `Planck slop' --- {\em i.e.} $1/M_p$ suppressed higher-dimension effective interactions, we identify another issue of 
UV sensitivity.\footnote{Here $M_{p} = 2.44 \times 10^{18} \, {\rm GeV}$ is the reduced Planck mass. We note that sensitivity to
`Planck slop' is also called the $\eta$-problem in some literature.}

The new issue we discuss first arises for inflationary models that are predictive in the sense that HI models are: that is, there are fewer free parameters than there are inflationary observables. In this case, the fact that the Renormalization Group (RG) running {\em even at low energies} required to relate inflationary predictions to other observables is UV sensitive also introduces new parameters into the predictions for quantities like $n_s$ and $r$.\footnote{The effect of UV physics, classified in terms of higher dimensional operators.
modifying the running of the SM parameters was recently completely calculated for the first time in Ref.\cite{Jenkins:2013zja}, for dimension six operators. We will use these results extensively in this paper in this application to cosmology.}

\subsection{UV Issues}

We here briefly describe in more detail, and contrast, the various kinds of UV sensitivity that can arise, in order to set the context for the quantitative calculation in the next sections of their effects in the minimal HI model.

An EFT analysis of inflation leads to the well-known observation that UV physics generically tends to modify the inflaton potential so strongly that it ruins the flatness that is responsible for the slow roll. It typically does so because integrating out UV physics at a scale $M$ contributes to low-dimension operators in the EFT --- like corrections to the vacuum energy or scalar masses, $\delta \cL = - \sqrt{-g} \left( c_0 + c_2 \, \phi^2\right)$ --- that are generically large: $c_0 \propto M^4$ and $c_2 \propto M^2$. This is the inflationary version of the standard `naturalness' problems that make it challenging to have light scalars within a generic EFT.

On the other hand, it is also known that once the low-dimension interactions are under control, UV physics can decouple from generic inflationary predictions \cite{Kaloper:2002cs, Burgess:2003zw}, just like it does from other types of low-energy phenomena (provided the UV physics is adiabatic \cite{Danielsson:2002kx, Martin:2000xs,Burgess:2002ub}). This is because corrections to high-dimension interactions are {\em suppressed}, rather than enhanced, by the large scale. If $\delta \cL = - c_k \, \sqrt{-g} \; \phi^k$ then $c_k \propto M^{4-k}$, which is suppressed for large $M$ if $k >4$. 
Of course the effective interactions satisfying $k \le 4$ can still be problematic.\footnote{In general the coefficient, $c_\ssQ$, of an operator $Q$ in $\delta \cL$ varies as $M^{4-d_\ssQ}$, where $d_\ssQ$ is the full scaling dimension of $Q$ (including anomalous dimensions). Note that a sensitivity of large-field models to higher-dimensional Planck slop can be due to large field excursions generating large anomalous dimensions for operators that were initially suppressed, potentially spoiling inflation as it progresses. See however Ref.\cite{Silverstein:2008sg} for a construction that avoids this.}

However even if such UV contributions are small in absolute size, they can still be large enough to ruin (or strongly perturb) inflation, since inflation requires not just that the inflaton mass be smaller than $M$; it must also be smaller than the Hubble scale, $\cH \sim V/M_p^2 \ll M$. Because of this, interactions suppressed by powers of $1/M$ can still contribute non-negligibly to slow-roll parameters --- and so also to $r$ and $n_s$ --- even if they do not ruin inflation. For instance, a $c_6 \, \phi^6$ contribution to the potential competes with an $m^2 \phi^2$ term whenever $c_6 \, \phi^4 \propto \phi^4/M^2 \simeq m^2 \lsim \cH^2$. This can actually happen (even if $M \simeq M_p$) because $\cH$ is itself Planck-suppressed relative to the other scales in the potential. For most inflationary models, however, the slow-roll parameters are not predicted in terms of other observables, so the standard approach simply rolls all such UV contributions into the uncertainty in the values of the slow-roll parameters, allowing them to be ignored in practice.

Our focus in this paper is on a third way UV physics affects the low-energy inflationary model, distinct from the above two well-understood issues. It first arises when the inflationary model involves fewer parameters than there are inflationary observables, such as in Higgs Inflation. For the HI model, one measures the couplings within the scalar potential in particle physics experiments at comparatively low energy, and inflationary predictions are then made in terms of these parameters. This raises a technical complication because the field values, $(H^\dagger H)_{\rm inf} \sim M_p^2/\xi$, associated with inflation are enormous relative to those, $(H^\dagger H)_{\rm vac} \sim v^2 \simeq (246 \; \hbox{GeV})^2$, relevant to particle physics. The extrapolation of the potential to fields this large involves large logarithms, whose leading behaviour can be summed using Renormalization Group (RG) methods. This RG-based extrapolation is an important step when relating the large-field/high-energy inflationary potential to the small-field/low-energy parameters inferred from particle physics measurements at electroweak (EW) energies \cite{DeSimone:2008ei, Bezrukov:2008ej, Bezrukov:2009db}. 

Our main point is that this RG improvement of the potential is also sensitive to the details of a host of higher-dimension effective interactions, most of which are {\em not}
pure Higgs field operators. For instance, by contrast with $\delta \cL = - c_6 \sqrt{-g} \; \phi^6$, an effective interaction like $\delta \cL = -\frac14 \, c_g \sqrt{-g}\; (H^\dagger H) F_{\mu\nu} F^{\mu\nu}$ does not contribute at tree level to the scalar potential, because of the presence of the gauge fields. However it {\em does} contribute at the quantum level because this operator contributes to the running of the corresponding gauge coupling of order $\delta (1/g^2) \sim c_g \, m_h^2/16\pi^2$ once the quantum fluctuations of the Higgs fields are calculated at one loop. This modification in the running of $1/g^2$ also feeds into the RG evolution of the other SM couplings at two loops, and this contributes to the running of the Higgs coupling, $\lambda$ \cite{Jenkins:2013zja}. As a consequence the value of the coupling $c_g$ can find its way into inflationary predictions.
%%yyy: I took this out because it is too detailed for the introduction: This latter effect is not as intuitive, but results from the use of the EOM in renormalizing the theory \cite{Jenkins:2013zja}.

Naively the size of any such contributions to $\delta\lambda$ would be expected to be very small. After all, if the effective interaction arises from integrating out a particle at mass $M$, then $c_g \, m_h^2 \propto m_h^2/M^2$, rapidly becomes very small for $M \gg m_h \simeq 125$ GeV.  Further, the specific example mentioned in the previous paragraph is  a two loop effect. However, there are also one loop effects of this form. Further, in the HI model, because inflation takes place at large values of the Higgs field, $H \sim M_p/\sqrt\xi$, and $m_h$ is itself proportional to $H$, the effective Higgs mass can also be very large. Restricting to the contributions of operators of the form $\delta \cL \sim H^2 F^2$, one finds a contribution to the running of $\lambda$ {\em at one loop} \cite{Jenkins:2013zja}
\be \label{d1grun}
 \delta \left( \mu \frac{d \lambda}{d \mu}\right) \supset \frac{m_h^2}{16\pi^2}\left[ 9 \, g_2^2 \, C_{\ssH \ssW} + 3 \, g_1^2 \, C_{\ssH \ssB}  + 3 g_1 \, g_2 \, C_{\ssH\ssW\ssB} \right] \sim \frac{g^2 m_h^2}{16\pi^2 \, M^2} \,,
\ee
where the final relation indicates the order of magnitude with $g$ generically representing $g_1$ and $g_2$, the couplings
of the $\rm SU_\ssL(2) \times U_\ssY(1)$ EW gauge bosons. See Ref  \cite{Jenkins:2013zja} for details on the operator notation used here.
This correction need not be inordinately small if $m_h \sim M$ at the values of $H$ of interest, even if $M$ itself is very large.

There are principally two kinds of uncertainty in this kind of expression. The first is what value to use for the mass, $M$, of any new threshold. Part of the framework of minimal HI is the assumption that there are no new heavy particles beyond the SM between EW and inflationary energies, because any such thresholds generically could introduce effective couplings --- like $(H^\dagger H)^3/M^2$, for example --- whose appearance within the potential could disturb the dynamics enough to destroy the inflationary slow roll.

\begin{figure}[t]
  \centering
  \includegraphics[width=5in,height=3.5in]{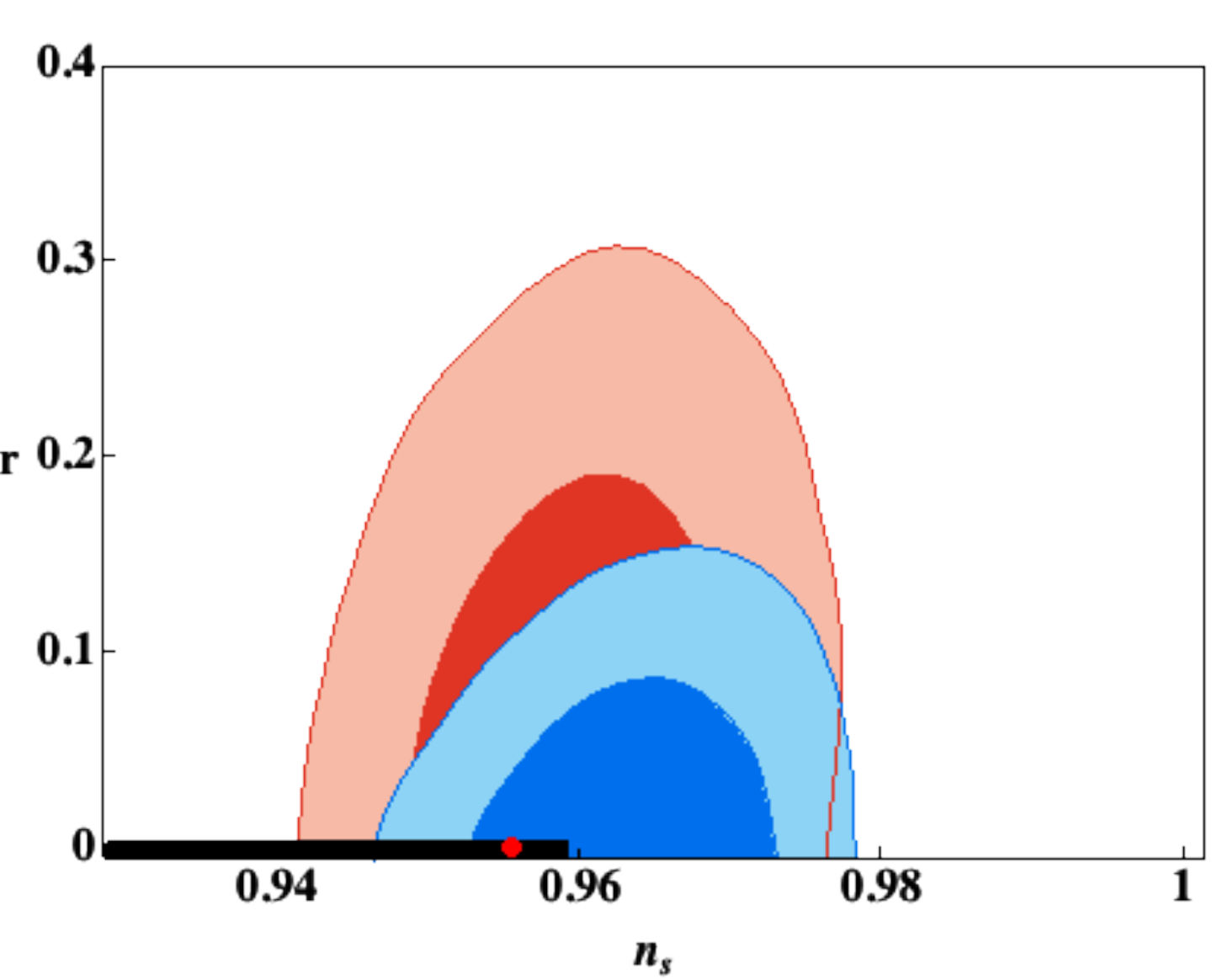}
  \caption{The potential effect of the unknown UV completion on the expectation for CMB parameters in the Higgs inflation scenario. The red dot
  is the prediction in HI for the scalar to tensor ratio $r$, and the spectral index $n_s$, without the effect of higher dimensional operators modifying the RG running. See Section \ref{nonlinear} for the details of how this prediction is obtained. The black line is the span of expected values for these parameters when the higher dimensional operators are also marginalized over. (The thickness of the line in the direction of $r$ is exaggerated so that the line is visible.) The figure also shows the one and two sigma regions of Fig 4 of Ref. \cite{Ade:2013uln}. The larger red regions are Planck and WMAP data + BAO + $\rm \Lambda CDM$ + r allowing running of $d n_s/d k$. The smaller blue regions are Planck and WMAP data + BAO + $\rm \Lambda CDM$ + r not allowing a running of $d n_s/d k$.}
  \label{Fig;nsvsr}
\end{figure}

However the mass, $M$, of the lowest new particle cannot be arbitrarily high. $M$ cannot be much larger than $ \sim \Lambda$, where $\Lambda$ is the `unitarity scale', or the upper limit of the domain of validity of the semi-classical approximation \cite{Burgess:2009ea}. $\Lambda = \Lambda(H^\dagger H)$ is Higgs-field dependent \cite{Bezrukov:2010jz}, and arises because the coupling to gravity is not renormalizable, and so the size of quantum effects can only be quantified within an EFT framework. Within this framework (as we review below) $\Lambda \sim M_p/\xi$ for the small fields, $H \ll M_p/\xi$, relevant to particle physics; while $\Lambda \sim M_p/\sqrt\xi$ for the larger fields, $H \simeq M_p/\sqrt \xi$, relevant to inflation. If we conservatively use $M \sim \Lambda \propto M_p/\sqrt\xi \sim H$ in the inflationary regime, and that $m_h \propto H$ there, we see that the correction in Eq.~\pref{d1grun} can be comparable to one-loop contributions computed within the SM. Effects such as this can be large enough to visibly change the implications for HI in the $n_s - r$ plane, as is illustrated in Fig.~\ref{Fig;nsvsr}.

%%yyy if you want to remove the figure, just uncomment the \ignore and \endignore instead of commenting out the whole section..

%\ignore
%\endignore

The other uncertainty in these estimates is whether or not all other higher dimensional operators in the non-renormalizable EFT actually vanish. They do not for any known proposals for weakly coupled physics beyond $\Lambda$ (such as string theory, or higher-dimensional gravity, for example). 
All of the higher dimensional operators will be generated by renormalization, so any vanishing of all these terms, if accomplished, will necessarily only occur at one scale. Further, one need not consider this question to be an exotic one purely in the context of gravity. We discuss in Section \ref{unitaritysection} how attempts to banish these operators can be mapped to analogous statements on unitarity violation involving massive spin one states in an EFT, with no need to invoke gravity. Attempts to argue away these operators would in this manner have broader implications for our understanding of unitarity violation and renormalization in many EFTs.

But these strong arguments, and the  absence of examples, does not remove the logical possibility that such physics might exist; sufficiently suppressing {\em all} dangerous dimension-six interactions at the scale $\Lambda$. It is difficult to say more without a specific and precise proposal for what the UV physics is that must enter at scale $\Lambda$,what the coefficients of the
operators will be\footnote{See for instance Refs. \cite{Antoniadis:2009rn, Antoniadis:2010nb} for a classification of the operators that might appear in the context of the MSSM. See also Ref. \cite{Chatterjee:2011qr} for a study of Higgs inflation embedded in the MSSM. We also note that some simple models have also been proposed to UV complete Higgs inflation and avoid the unitarity bound,see for example \cite{Giudice:2010ka}.}. Given our current lack of knowledge of physics beyond the Standard Model, our own point of view is that these unknown order-unity coefficients are likely to be nonzero and so represent intrinsic theoretical uncertainties that must be propagated through to low energy observable quantities in EFT's, such as CMB observables in HI\footnote{In a similar spirit, see \cite{Branchina:2013jra} for a discussion of the sensitivity of the stability of the EW vacuum to new physics.}.

The importance of these threshold-like terms within RG equations was recently emphasized for non-cosmological applications in Ref.\cite{Jenkins:2013zja}, as part of a systematic renormalization program of the SM EFT (with full flavour structure) completely carried out in Refs. \cite{Grojean:2013kd,Jenkins:2013zja,Jenkins:2013wua,Alonso:2013hga}. In Ref.\cite{Jenkins:2013zja} the complete modification of the running of the parameters present in the renormalizable SM Lagrangian due to dimension six operators was explicitly calculated.  In what follows we use these results to illustrate how the running of SM parameters can be modified in the case of HI. We discuss how this impacts attempts to predict $n_s$ and $r$ in this model, and derive the results illustrated in Fig.~\ref{Fig;nsvsr}.

Our broader lesson is this: although we discuss in detail HI, similar issues should arise within the SM RG in any attempts to link EW scale physics with the higher scales involved in inflationary (and other cosmological) scenarios\footnote{Any complete account of reheating into some sector that contains the Standard Model seems to force this issue upon us by directly or indirectly coupling the SM degrees of freedom to the inflaton, for example.}. UV sensitivity is a many-headed hydra, and it is only with the development of more predictive models that this latest version has become potentially relevant.

The outline of this paper is as follows, in Section II we discuss HI and the cut off scales present in theories of this form. In Section III, we discuss the RG evolution used in these theories, and we outline the contributions to the RG equations that we include that were previously neglected. We then demonstrate how these corrections impact predictions in these theories based on EW scale measurements. Finally, in Section IV, we conclude.

%%%%%%%%%%%%%%%%%%%%%%%
\section{Higgs Inflation and UV Physics}\label{cutoffs}

In this Section we will review the Higgs inflation framework; discuss some of the issues that arise from its treatment within an EFT framework; and present how the HI gravity-Higgs mixing modifies the RG evolution of effective operators within the EFT.

\subsection{The model}

The HI model \cite{Bezrukov:2007ep} proposes to use the SM Higgs field as a single-field inflaton, with the Higgs playing the (particularly economical) role of a  non-minimally coupled inflaton, along the lines studied in \cite{Spokoiny:1984bd, Salopek:1988qh}. The theory is defined by the Lagrangian density
%
%yyy Took the extra operators out here, but will put them back once the EFT discussion begins below...
%
\bea \label{HIlagr}
 \mathcal{L}_\HI = \mathcal{L}_{\SM} - \sqrt{- \hat g} \left[ \frac{M_p^2 }{2}  +
 \xi \, (H^\dagger H ) \right] \hat R  \,,
\eea
where $\cL_\SM$ is the usual Standard Model Lagrangian density with the flat metric replace by a general `Jordan-frame' metric, $\hat g_{\mu\nu}$, whose Ricci scalar is denoted $\hat R$.

The idea is to use the SM Higgs as the inflaton, and because the SM potential is not particularly flat the inflationary slow roll is sought at large Higgs field values. This turns out to be possible when $H \sim M_p/\sqrt{\xi}$. Primordial fluctuations are then assumed to be generated from quantum fluctuations in the usual way, and their amplitude can be made to agree with CMB observations by choosing $\xi \simeq 10^4$.\footnote{Recent versions of Higgs inflation tune the top and Higgs mass and consider much smaller $\xi \sim 10$. See the comments at the end of the paper.}

The theory is easiest to analyze in the Einstein frame, with the metric canonically normalized. To do so use the Weyl transformation $\hat g_{\mu\nu} \rightarrow g_{\mu\nu}$ given by
\bea
 \hat g_{\mu\nu} = f \, g_{\mu\nu}   \quad \quad \hbox{with} \quad \quad f = \left[1 + 2 \, \xi ( H^\dagger H) /M_p^2 \right]^{-1} \,.
\eea
After making this replacement the terms of particular interest in HI are given by
\bea
 \frac{ \mathcal{L}_\HI}{\sqrt{-g}} &=& - \frac{1}{2}
 \, M_p^2 \, R  - V_\EF(H^\dagger H) - g^{\mu \nu}
 \left[ f (D_\mu H)^\dagger \, (D_\nu H)
  + \frac{3 \, \xi^2 f^2}{M_p^2} \,
 \partial_\mu (H^\dagger \, H) \,
 \partial_\nu(H^\dagger \, H) \right], \nonumber
\eea
where $R$ is the Einstein-frame Ricci scalar built using $g_{\mu\nu}$, and the Einstein-frame Higgs potential is
\be \label{EFhpot}
 V_\EF = f^2 \, V_\SM = \lambda f^2 \left( H^\dagger H - \frac{v^2}{2} \right)^2 \,.
\ee
HI exploits the fact that $f \propto (H^\dagger H)^{-1}$ for large enough expectation value of $H^\dagger H$, and so because $V_\SM \propto (H^\dagger H)^2$ for large $H^\dagger H$, $V_\EF$ becomes flat enough to inflate in the large-field regime. More quantitatively the potential flattens once $H^\dagger H \gsim M_p^2/\xi \gg v^2$ and so this defines the inflationary regime.

It is most efficient to move to unitary gauge, $\sqrt2 \, H = (0,v+h)^{T}$ and then perform the field redefinition $h \to \chi(h)$ that puts the scalar kinetic energy into canonical form: $- \frac12 \, \sqrt{-g} \; g^{\mu\nu} \partial_\mu \chi \partial_\nu \chi$. The required redefinition satisfies
\bea\label{fieldredefinition}
 \frac{d \, \chi}{d \, h} =  \frac{\left[1+ (\xi +  6 \, \xi^2) \,(h/M_p)^2\right]^{1/2}}{1 + \xi \,
 (h/M_p)^2} \,,
\eea
which for large $\xi$ is easily integrated. In the small-field regime, where both $h$ and $\chi$ are much smaller than $M_p/\xi$, it integrates to
\be \label{hvschismall}
 h \simeq \chi - \frac{\xi^2 \chi^3}{M_p^2} + \cdots \qquad \hbox{(when $h,\; \chi \ll M_p/\xi$ and $\xi \gg 1$)}\,;
\ee
and in the large-field regime, $h \gg M_p/\xi$, we instead find
\be \label{hvschilarge}
 h^2 \simeq \frac{M_p^2}{\xi} \Bigl( e^{\beta \chi} - 1 \Bigr)
 \qquad \hbox{(when $h \gg M_p/\xi$, and so $\beta \chi \gg \cO(1/\xi)$)}\,,
\ee
where the parameter in the exponent is
\be
 \beta = \frac{1}{M_p} \sqrt{\frac23}  \,.
\ee
In both cases we choose integration constants to ensure $h=0$ corresponds to $\chi = 0$.

It is the large-field form of the potential that is relevant to inflation,
\bea \label{HIpot}
 V_\EF(\chi) \simeq \frac{\lambda \, M_p^4}{4 \, \xi^2} \Bigl( 1 - e^{-\beta \chi} \Bigr)^2 \,,
\eea
which is exponentially flat deep within the large-field region. For cosmological applications this translates into the following $\chi$-dependent Hubble scale and slow-roll parameters,
\be
 \cH^2 \simeq \frac{\lambda M_p^2}{12 \, \xi^2} \Bigl( 1 - e^{-\beta \chi} \Bigr)^2 \,, \qquad
 \epsilon \simeq \frac43 \left( \frac{ 1}{e^{\beta \chi}-1} \right)^2 \,, \qquad
 \eta \simeq - \frac43  \left[ \frac{ e^{\beta \chi} - 2 }{(e^{\beta \chi} - 1)^2} \right] \,,
\ee
in terms of which the spectral index, $n_s$, and the tensor to scalar ratio, $r$, are given by the standard formulae \cite{Liddle:1992wi},
\bea
 n_s = 1 - 6 \, \epsilon_* + 2 \, \eta_* \qquad \hbox{and} \qquad r = 16 \, \epsilon_* \,,
\eea
where the subscript `*' indicates evaluation at the epoch of horizon exit. Inflation ends when $\beta \chi \simeq \cO(1)$, when the slow-roll parameters are not small, if one assumes $N_e \simeq 57.7 \pm 0.2$ $e$-folds of inflation \cite{Bezrukov:2013fka} then $\beta \chi_* \simeq 4$ at horizon exit, giving the successful predictions $n_s \simeq 0.967$ and $r \simeq 0.0031$, which, at leading order depends only on the SM parameters through the Higgs self-coupling, $\lambda$ (other than the implicit dependence on SM parameters in the reheating analysis that is used to fix the value assumed for $N_e$).

\subsection{Embedding into an EFT}

Because the HI model includes gravity its semiclassical expansion is not renormalizable, even though the coupling $\xi$ is dimensionless. As such, the only known way to systematically calculate its quantum properties is to interpret it as an EFT, regarding Eq.~\pref{HIlagr} as the leading terms in a low-energy expansion (see, for example, \cite{Donoghue:1995cz,Burgess:2003jk} for an introduction within a gravitational context),
\bea
 \mathcal{L}_\EFT = \mathcal{L}_{SM} - \sqrt{- \hat g} \left[ \frac{M_p^2 }{2}  +
 \xi \, (H^\dagger H ) \right] \hat R  + \sqrt{- \hat g} \sum_i C_i \, Q_i  \,,
\eea
where the operators $Q_i$ consist of all possible interactions built from the given fields consistent with the low-energy gauge symmetries. Their effective couplings, or Wilson coefficients, $C_i$, are generically suppressed by powers of the large scale,
$M$, of the massive states that were integrated out to generate $\cL$ in the first place. 

The scale $M$ need {\em not} be $M_p$. Generally it is the smallest mass scale appearing in a denominator that usually dominates.
Further, the scales suppressing different fields, or derivatives, need not coincide in general, see Refs. \cite{Manohar:1983md,Jenkins:2013sda,Buchalla:2013eza} for some discussion on power counting. In the discussion that follows, for simplicity, we will assume that the suppression scale is generically $M$.
Also note that curvature-squared terms need not be suppressed by $M$, but in four dimensions curvature-squared terms can be eliminated using an appropriate field redefinition, and so are redundant interactions.

There are an infinite number of potential operators, $Q_i$, but only a finite number that are suppressed by less than a specific power of $1/M$. It is by organizing calculations in powers of $1/M$ that calculations become predictive, if only finite accuracy is demanded. Terms in $\cL$ involving the fewest powers of $1/M$ are expected to dominate at low energies if $M$ is very large.

EFT makes two of the choices made by HI appear very natural. First, part of what is attractive about the HI model is that its only new interaction has engineering dimension of (Energy)${}^4$, and so its coupling is unsuppressed by $1/M$. Furthermore, it is the only such term possible that involves SM fields and that is not already included within $\cL_\SM$. This is attractive because such terms might plausibly dominate in the EFT at low energies when $M$ is large.

Second, the $1/M$ expansion has two logically distinct parts: expansions in powers of derivatives; and expansions in powers of fields (like $H$). 
An EFT  reproduces the same S matrix elements as the full theory in some momentum regime of validity.
Although this requires derivatives be small, it need not also require small fields, unless the scalar potential is such that large fields also imply large energy. For potentials like Eq.~\pref{HIpot},  large fields do not imply large energies and so nothing in the EFT {\em a-priori} requires $h$ be small, even in comparison with $M_p$.

On the other hand, nothing seems to require that large values of $h$ must correspond to low energies, and so it would be natural to expect the sum over $Q_i$ to include terms like $(H^\dagger H)^n/M^{2(n-2)}$, with $n > 2$, or $(H^\dagger H)^n R/M^{2(n-1)}$, with $n > 1$. Such operators would be dangerous for inflation to the extent that they ruined the property that $f^2 V_\SM$ becomes constant at large fields.

In HI such higher powers of $H^\dagger H$ are assumed {\em not} to arise, and this is an implicit condition on the kinds of UV completion for which $\cL_\HI$ can be the low-energy limit. One way this might happen is if no heavy particles were present at all with masses below the fields needed for inflation, such as if the smallest such UV mass satisfies $M \gg M_p/\sqrt\xi$. Alternatively one might hope for some sort of strong UV dynamics that provides an anomalous dimension for $H^\dagger H$ that suppresses the dangerous terms more than they would naively be. Or one can hope the UV theory has a symmetry, like scale invariance, that can suppress such terms. Unfortunately, to our knowledge, no precisely defined candidate theory exists that accomplishes any of these hopes in detail.

Further, integrating out heavy particles also normally contributes corrections to the Higgs mass that are $\delta m_h^2 \sim M^2$; the usual EW hierarchy problem. Since this only becomes a problem once a heavy particle is integrated out, this problem can also be pushed up to very high energies if it is assumed that no new particles exist beyond the SM at lower energies. HI assumes (as do most other inflationary models) that somehow the unknown UV physics does not generate these dangerous effective interactions when integrated out. For the purposes of our later arguments, we follow suit in the rest of this paper and assume the required type of UV physics exists.

In the next sections we describe another way that UV physics can complicate the low-energy inflationary story, where we focus on a different set of operators, $Q_i$. We consider in detail the subset of dimension six operators constructed purely of the SM field content and consistent with (linearly realized) $\rm SU_c(3) \times SU_L(2) \times U_Y(1)$ gauge invariance. The list of possible dimension six operators has been known for some time \cite{Buchmuller:1985jz}, and the minimal basis with redundant operators eliminated using lower-order field equations --- or, equivalently, using appropriate field redefinitions is now known\footnote{There are 59 operators neglecting flavour indicies, or 2499 unknown parameters characterizing beyond the SM physics in this case, when flavour indicies are not neglected \cite{Alonso:2013hga}.} \cite{Grzadkowski:2010es}. We use this operator basis in what follows to characterize how the unknown UV completion can effect the running of the SM parameters below the scale $\Lambda$.

Because the size of these (and other) operators are controlled by $1/M$, we first 
pause to review the argument that there is an upper bound to how big the mass, $M$, of the UV threshold can be.

\subsection{The unitarity scale and the nonlinear realization}\label{unitaritynonlinear}

The only known systematic way to incorporate quantum effects in non-renormalizable field theories is to interpret them as an EFT, within an implicit low-energy expansion. If mistakenly this expansion is used at too high an energy, the low-energy expansion breaks down, leading to a loss of predictiveness. This problem is often cast in terms of unitarity violation,\footnote{Typically the Hamiltonian constructed from a real Lagrangian density is Hermitian, and if so the theory must be unitary. Yet unitarity is inconsistent with cross sections that rise too quickly with energy, so if such a cross section is found it implies an approximation has failed in the derivation. The offending approximation is usually the low-energy approximation implicit in using the non-renormalizable theory in the first place.} with the scale, $\Lambda$, above which the low-energy theory fails called the unitarity scale. 

For HI the scale $\Lambda$ is of interest because it provides an upper limit to the energy range over which the theory can apply without modification. As such it provides an upper bound on the mass scale, $M$, of the first new UV state not already contained within HI itself. 

\subsubsection*{The HI unitarity scale}\label{unitaritysection}

Because the coupling $\xi$ is large, it exacerbates the breakdown of the low-energy approximation, and as a result lowers $\Lambda$ relative to its naive value, $M_p$, associated with pure gravity. It does so in a way that depends on the size of the background Higgs field \cite{Bezrukov:2010jz} with
\be
 \Lambda \simeq \frac{M_p}{\xi} \quad \hbox{when} \quad h \lsim \frac{M_p}{\xi} \,,
 \qquad \hbox{and} \qquad
 \Lambda \simeq \frac{M_p}{\sqrt\xi} \quad \hbox{when} \quad h \gsim \frac{M_p}{\sqrt\xi} \,. \label{two} 
\ee
Why these results are obtained will be reviewed in detail below. The cut off scale depends on the channel considered, see Ref \cite{Ren:2014sya}
for a recent discussion of various channel cut off scales.
The overall cut off scale quoted for the effective theory depends on the lowest cut off scale found. The low-field value for $\Lambda$ was determined in Ref.~\cite{Burgess:2009ea} by using powercounting, to systematically identify the lowest cut off scale present. In our detailed numerics we use the lowest cut off scale given by Eqn \ref{lamdep}. The cut off scale can be easily discovered in the theory in some particular cases. Expanding the $\xi (H^\dagger H) \hat R$ term about Minkowski space, using $\hat g_{\mu \, \nu} = \eta_{\mu\nu} + h_{\mu \, \nu}/{M_p}$ and tracking the metric-scalar mixing in the Jordan frame, we have
\bea\label{jordanmixing}
 - \sqrt{- \hat g} \; \xi (H^\dagger H) \,\hat R \simeq \frac{\xi}{M_p} \, h^2 \, \eta^{\mu\nu} \, \partial^2 \, h_{\mu \nu} \, \, + \cdots ,
\eea
showing the explicit dependence on the scale $M_p/\xi$. This scale was also shown to be present in the explicit expansion of the potential \cite{Barbon:2009ya}
at small field values.\footnote{See Ref. \cite{Han:2004wt} for scattering results that support this point.} Further, this scale is also found in any gauge (including unitary gauge), and in both the Jordan and Einstein frames \cite{Burgess:2010zq}, when calculating in the EW vacuum. Note that the cut off scale being the same in the Jordan and Einstein frames, and in unitary gauge, is in conflict with some claims in the literature, see however \cite{Burgess:2010zq} for clarifications on both of these points.

Once $h$ climbs above $M_p/\xi$ the scale $\Lambda$ also climbs due to the suppression of the physical Higgs interactions due to its mixing with the metric in Eq.~\pref{jordanmixing} \cite{Bezrukov:2009db, Bezrukov:2013fka}.  It is because $\Lambda$ rises to $M_p/\sqrt\xi$ within the inflationary regime that it can be consistent to consider Hubble scales as large as $\cH \sim M_p/\xi$ without invalidating the semiclassical approximation \cite{Bezrukov:2010jz}.

\subsubsection*{The nonlinear realization}\label{nonlinearsection}

The SM Higgs couplings are the unique ones that allow unitarity to be valid at scales far above the Higgs vev, $\bar{h}$, in the presence of massive spin one states whose mass is generated by the scale $\bar{h}$. Once the Higgs couplings become modified (as they are by mixing with the metric) there is generically a unitarity problem at scales of order $4 \, \pi \, \bar{h} \, c$. Here $c$ is a schematic coefficient 
that indicates the degree of deviations in the effective Higgs couplings from the SM values.
An example in Appendix \ref{HiggsAx} gives an illustrative toy description of this mixing, and
gives some intuition for why the problems at the scale $M_p/\xi$ do not dominate once $4\pi \bar{h} c$ becomes the larger scale of the two unitarity limits in the peculiar case of HI. 

It can be convenient not to use unitary gauge and instead to rewrite the theory to display explicitly the would-be Goldstone bosons of EW symmetry breaking, and how these interact with the scalar Higgs singlet \cite{Bezrukov:2009db, Bezrukov:2013fka}. For later convenience we summarize these couplings here, and show how they also can be used to infer the size of $\Lambda$ in different regimes. Consider a general EFT with a nonlinearly realized $\rm SU(2) \times U(1)$ in the scalar sector, massive vector bosons due to a classical background field vev, and a scalar singlet with general couplings.\footnote{For an introduction to the concept of a nonlinearly realized symmetry, see Ref. \cite{Weinberg:1968de}.} In recent years, this EFT formalism is under intense development as an alternative EFT description of the observed Boson at LHC, see Refs.~\cite{Grinstein:2007iv, Contino:2010mh,Azatov:2012bz, Alonso:2012px, Buchalla:2012qq, Buchalla:2013rka}. (See \cite{Burgess:1999ha} for a similar unitary gauge formulation of Higgs properties.) We write the theory in the frame where the scalar field and graviton have been canonically normalized {\em i.e.} using $\chi$ in the Einstein frame, and the Goldstone bosons eaten by the $W^\pm$ and $Z$ bosons are denoted by $\pi^a$ where $a = 1,2,3$. The Goldstones are grouped together as
\bea
 \Sigma(x) = e^{i \sigma_a \, \pi^a/\bar{\chi}} \; ,
\eea
with $\bar{\chi}$ the background $\chi$ vev. The $\Sigma(x)$ field transforms
linearly under $\rm SU(2)_L \times SU(2)_R$ as $\Sigma(x) \rightarrow L \, \Sigma(x) \, R^\dagger$ where $L,R$ indicate the transformation on the left and right under these groups. The diagonal subgroup of $\rm SU(2)_L \times SU(2)_R$ is called the `custodial' group, and the physical Higgs, $\chi$, is a singlet under this group. 

The leading terms in a derivative expansion are given by 
\be\label{nonlinear}
 \frac{\mathcal{L}_\HI}{\sqrt{-g}} = -\frac{1}{2} (\partial_\mu \chi)^2 - V(\chi) - \frac12 \, F^2(\chi) {\rm Tr} (D_\mu \Sigma^\dagger \, D^\mu \Sigma) 
  - \frac{1}{\sqrt{2}} \, (\bar{u}_L^i \bar{d}_L^i) \, \Sigma \, 
 \left(
 \begin{array}{c}
 y_{ij}^u \, u_R^j \\
 y_{ij}^d \, d_R^j
 \end{array} \right) Y(\chi) +  h.c. \,,
\ee
where the potential is given by Eq.~\pref{EFhpot}, or equivalently Eq.~\pref{HIpot} in the large-field limit. Similarly the functions $F^2$ and $Y$ are given by
\be
 F^2(\chi) = \frac12 \, f \, [v + h(\chi)]^2 \qquad \hbox{and} \qquad
 Y(\chi) = f^{1/2} [v + h(\chi)] \,.
\ee
\subsubsection*{Scattering amplitudes}

A virtue of explicitly using a chiral EW lagrangian is that one can make direct contact with many previously obtained results in the literature, and frame questions
about unitarity violation in HI, in terms of equivalent questions and claims for the scattering of massive spin one vectors. For example, arguments
that higher dimensional operators will not be present suppressed by the scale $\sim M_p/\xi$ are related, in this formalism, to claims
about solving the unitarity violation problems of the SM, with no Higgs particle, and no higher dimensional operators. The later physics is more familiar
to many, so this can be advantageous. As Gravity is then no longer essential to the discussion, this has the potential
to clarify claims in the literature about the nature of unitarity violation in HI, and possible solutions to this problem. An example is the scattering amplitudes for particles computed in a semiclassical expansion around the classical background field,
\be
 \chi = \bar{\chi} +  \hat{\chi}.
\ee
Strictly speaking this scattering is normally computed when $\bar\chi = 0$ takes its vacuum value, but it can also be done for more general $\bar\chi$, even if these are not at extrema of the classical potential. Scattering can be computed provided the quanta involved are energetic enough that the background evolution is effectively adiabatic. In much the same way that we compute scattering in the present epoch despite the overall cosmological expansion of the universe.

To this end we expand $F^2$, $Y$ and $V$ as follows
\bea\label{nonlinearfns}
  F^2(\bar \chi + \hat \chi) &=& \bar{\chi}^2 \left[1 + 2 \, a \, \frac{\hat{\chi}}{\bar{\chi}} + b \, \frac{\hat{\chi}^2}{\bar{\chi}^2}  +  b_3 \,  \frac{\hat{\chi}^3}{\bar{\chi}^3} + \cdots \right], \nn\\
 Y(\bar \chi + \hat \chi) &=& \bar{\chi} \left[1 + c \,  \frac{\hat{\chi}}{\bar{\chi}}  +  c_2 \,  \frac{\hat{\chi}^2}{\bar{\chi}^2}  + \cdots \right] \,,
\eea
and
\be
 V(\hat{\chi}) = \frac{1}{2} \, m_\chi^2 \, \hat{\chi}^2 + \frac{d_3}{6} \, \left(\frac{3
    \, m_\chi^2}{\bar{\chi}} \right) \, \hat{\chi}^3 + \frac{d_4}{24} \left(\frac{3
    m_\chi^2}{\bar{\chi}^2} \right) \hat{\chi}^4 + \cdots \; .
\ee
Here we use the notation of Ref \cite{Contino:2010mh}, suitably modified.
%%yyy dropped the discussion of the minimal coupling controversy since it probably is best not to mix that fight in with this one :)
%
Scattering in theories of this form was reveiwed, for example, in Ref.\cite{Contino:2010rs} (and references therein). 
The parameters in Eqn. \ref{nonlinear} 
in the SM, with no non-minimally coupled term, are $\left(a,b,b_3,c,c_2\right)_{sm} = \left(1,1,0,1,0 \right)$.
The scattering of the would-be goldstone bosons is given in terms of these parameters by
\be\label{goldstoneamplitude}
 \mathcal{A}(\sigma^i \, \sigma^j \rightarrow \sigma^k \, \sigma^l) = \left(1- a^2 \right) \left[ \frac{s \, \delta^{ij} \, \delta^{kl} + t \, \delta^{ik} \, \delta^{jl} + u \, \delta^{il} \, \delta^{jk}}{\bar{\chi}^2} \right] \,,
\ee
where $s$, $t$ and $u$ are the Mandelstam variables. Scattering into fermion final states (generally denoted $\psi$) similarly go as
\bea \label{goldstonepairprod}
 \mathcal{A}(\sigma^i \, \sigma^j \rightarrow \bar{\psi} \, \psi) &=&   \delta^{ij} \,  \frac{y_\psi \, \sqrt{s}}{\bar{\chi}} (1 -  a \, c ).
\eea
Using the chiral EW Lagrangian formalism, we can apply these results directly to the small- and large-field limits of $\cL_\HI$, we can thereby read off the scale $\Lambda$ in these limits.  

\medskip\noindent{\em Small-field limit}

\medskip\noindent
Specializing to the small-field form for $h(\chi)$, the shifts in these chiral EW parameters, due to the non-minimally coupled gravitational interaction, at low field values are
\bea
\delta \left(a,b,b_3,c,c_2\right) = - \frac{\xi^2 \, \bar{\chi}^2}{M_{pl}^2} 
\left(1, - \frac{12 \, \xi \, \bar{\chi}^2}{M_p^2} + \frac{6 \, \bar{\chi}^2}{\xi \, M_p^2} , 2,-\frac{3 \, \xi^3 \bar{\chi}^4}{2 \, M_p^4} + 3 \xi \frac{\bar{\chi}^2}{M_p^2} - \frac{3}{2 \, \xi},-\frac{3 \, \xi \bar{\chi}^2}{2 \,M_p^2}  + \frac{3}{2 \, \xi} \right). \nonumber
\eea
In the large $\xi$ limit, when considering field values around $M_p/\xi$ one can simplify this result to 
\bea
\delta \left(a,b,b_3,c,c_2\right) = - \frac{\xi^2 \, \bar{\chi}^2}{M_{pl}^2} 
\left(1, 0, 2,0, \right).
\eea
The Higgs mass is also redefined, as $m_\chi^2 \simeq 3 \, \lambda \, \bar{\chi}^2$.
The values of the couplings in the potential become $d_{3,4} \simeq 2/3$ in the large $\xi$ limit, for field values $\sim M_p/\xi$.
Now consider the effect of these modifications of the SM couplings. The Goldstone scattering is given by
\bea\label{goldstoneamplitude}
\mathcal{A}(\sigma^i \, \sigma^j \rightarrow \sigma^k \, \sigma^l) &=& \left(1- (a_{sm} + \delta a)^2 \right) \frac{s \, \delta^{ij} \, \delta^{kl} + t \, \delta^{ik} \, \delta^{jl} + u \, \delta^{il} \, 
\delta^{jk}}{\bar{\chi}^2}, \\
&=& \frac{2 \, \xi^2}{M_{pl}^2} \,  \left(s \, \delta^{ij} \, \delta^{kl} + t \, \delta^{ik} \, \delta^{jl} + u \, \delta^{il} \, \delta^{jk}\right) \nonumber,
\eea
in terms of the Mandelstam variables $s,t,u$. The scattering involving fermion fields, generally denoted $\psi$, and the singlet scalar go as
\bea
\mathcal{A}(\sigma^i \, \sigma^j \rightarrow \bar{\psi} \, \psi) &=&   \delta^{ij} \,  \frac{y_\psi \, \sqrt{s}}{\bar{\chi}} (1 - (a_{sm} + \delta a) \, (c + \delta c)), \\
&\simeq&  \frac{\xi^2}{M_{pl}^2} \, \delta^{ij} \, y_\psi \, \bar{\chi} \, \sqrt{s}.
\eea 

It is again established that the cut off scale in the EW vacuum is set by the scale $\Lambda_{ew} \simeq M_{pl}/\xi$. The
background field dependence cancels in an interesting manner in pure Goldstone scattering.
The independence of $\Lambda_{ew}$  on the background field value is due to the
modifications of the Higgs couplings being a perturbation $\propto  \bar{\chi}^2$. This is due to the fact that 
this modification is proportional to the background field value in the kinetic mixing of the singlet Higgs with the graviton. This
makes clearer why the scale of unitarity violation at low field values does not depend
on $\bar{\chi}$, contrary to the case of large field values.

Now consider the case where there is only a single scalar field that gets a vev, $S$, which generates a 
a massive vector through the Higgs mechanism.
It is known in explicit calculations of non-minimally coupled scalar fields to gravity, that in the case of a singlet scalar
field, some of the scattering amplitudes that lead to unitary violation in the case of multiple scalars, do not lead to unitarity violation \cite{Huggins:1987ea,Huggins:1987eb}.
The exact same conclusion is obtained in 
Eqn.\ref{goldstoneamplitude}, when all the Goldstone indicies coincide, as $i=j=k=l$, and the Mandelstam
relation on $s + t +u = \sum_i m_i^2$ cancels the high energy growth. This analogy has been noticed before, see Ref.\cite{Hertzberg:2010dc},
but the exactness of the correspondence is made clear with the non-linear chiral Lagrangian formalism.

As one approaches the scale $\Lambda_{ew}$, the arguments of Ref. \cite{Burgess:2009ea,Burgess:2010zq} establish
that the cut off scale remains at $\Lambda_{ew}$, although a small field perturbative expansion into the non-linear EW chiral Lagrangian begins to fail.

\medskip\noindent{\em Large-field limit}

\medskip\noindent
Switching to the large-field form for $h(\chi)$, Eq.~\pref{hvschilarge}, we read off parameter values $\bar \chi^2 \simeq M_p^2/\xi$ and
\be
 a = \frac{1}{\sqrt{6\xi}} \, e^{-\beta \bar \chi} \,, \quad
 b = - \frac{1}{3\xi} \, e^{- \beta \bar \chi} \,, \quad
 b_3 = \frac{1}{9\xi} \sqrt{\frac{2}{3\xi}} \; e^{-\beta \bar\chi} \,,
\ee
where we focus on the regime of inflationary interest where $e^{-\beta \bar\chi} \ll 1$. For these values $a \ll 1$ and so the rising cross sections of eqs.~\pref{goldstoneamplitude} become
\be\label{goldstoneamplitude}
 \mathcal{A}(\sigma^i \, \sigma^j \rightarrow \sigma^k \, \sigma^l) = \frac{\xi}{M_p^2} \Bigl[ s \, \delta^{ij} \, \delta^{kl} + t \, \delta^{ik} \, \delta^{jl} + u \, \delta^{il} \, \delta^{jk} \Bigr] \,,
\ee
showing that unitarity problems arise once energies reach the scale $s \sim \Lambda^2 \sim M_p^2/\xi$.
Between the scales $M_p/\xi$ and $M_p/\sqrt{\xi}$ the cut off scale rises as $\sim 4 \, \pi \bar{\chi}$, essentially as a theory with un-Higgsed massive spin one fields,
whose mass is set by the scale $\bar{\chi}$ \cite{Bezrukov:2009db, Bezrukov:2013fka}.

\medskip\noindent{\em Non-linearities in the SM}

\medskip\noindent

It is interesting to note that the physics discussed in the previous sections is clearly present in the SM, at least to some degree.
Even at low field values, once the Higgs gets a vev and breaks the $\rm SU(2) \times U(1)$ symmetry, a non-minimal gravitational coupling term leads to a non-canonical theory. Canonically normalizing reflects the symmetry breaking
back to a shift in the couplings of  singlet $\chi$, compared to the SM value. This effect can be incorporated by expressing the the EFT  as a non-linear realization
of  $\rm SU(2) \times U(1)$. So long as the Higgs gets a vev and the theory is written in curved space, a non-linear realization results, in the sense that the couplings of the canonically normalized scalar field deviate from the value expected in a linear realization of $\rm SU(2) \times U(1)$. This is true even when higher dimensional operators are allowed, as the $\rm SU_L(2)$ symmetry that relates these scalar couplings to the couplings of the eaten Goldstone Boson modes is broken.\footnote{See Ref.\cite{Isidori:2013cga,Brivio:2013pma} for some recent discussion on the differences between a linear and non-linear Higgs EFT.} 
Renormalizing the SM in curved space generates $H^\dagger \, H  R$ \cite{Birrell:1982ix}, so this physics is present in the SM in our spacetime.
The small corrections $\mathcal{O}(v^2/M_{p}^2)$ that introduce the non-linearity, due to the non-minimally coupled gravitational interaction, are implicitly always neglected when a linear EFT is used. This is manifestly a good approximation for almost all applications, but it is amusing to note
that the Higgs part of the SM EFT is always fundamentally non-linear in this manner. The main distinction in HI, is that one takes the expected coupling to not be of loop size, $ \sim 1/16 \pi^2 $, or of the order expected in a conformal theory, $1/6$, but instead $\xi \sim 10^4$, and studies the resulting theory at very large background field values.

%We also note that if the effective planck scale is low $M_p^{eff} \ll M_p$, as in some extra-dimensional scenarios, that attempt to naturally solve the hierarchy problem, this generic %effect of non-linearity due to kinetic mixing with the graviton can be more pronounced, even at low Higgs vev's. The naturalness of these scenarios is frequently tied to 
%the rough relation $v \sim M_p^{eff}$. Searching for hints of this non-linearity is also interesting in this regard. At the same time, this generic introduction of non-linearity indicates %that extra dimensional theories can have unitarity problems at the scale
%\bea
%\Lambda_{ED} \simeq \frac{M_p^{eff}}{\xi}
%\eeascan
%where $\xi$ in this case is the coupling that is present in the extra dimensional scenario, even when the Higgs mechanism is fully present
%to give mass to the vector bosons of the SM. Of course, the degree to which an actually scenario suffers from this potential issue depends upon detailed model building.
%However, the non-minimal coupling term {\em is} consistent with all the symmetries of the SM, and as such will generically be present when the theory is renormalized in curved %space in these scenarios, so $ \xi \gtrsim g_{SM}^2/16 \pi^2$ is expected.

\subsection{RG running in HI}\label{CWcorrections}

The one loop corrections to the usual
Coleman-Weinberg (CW) potential \cite{Coleman:1973jx} are incorporated in Higgs inflation as a perturbative correction to
$V_E$.  The leading corrections to the effective potential are
\bea\label{CWcorr}
\delta V &=& \frac{6 \, m_W(\bar{h})^4}{64 \, \pi^2} \, \left[\log \frac{m_W(\bar{h})^2}{\mu^2} - \frac{5}{6} \right] +  \frac{3 m_Z(\bar{h})^4}{64 \, \pi^2} \, \left[\log \frac{m_Z(\bar{h})^2}{\mu^2} - \frac{5}{6} \right]  \\
&\,& - \sum_f \frac{3 \, m_f(\bar{h})^4}{16 \, \pi^2} \, \left[\log \frac{m_f(\bar{h})^2}{\mu^2} - \frac{3}{2} \right], \nonumber
\eea
in the $\rm \overline{MS}$ scheme \cite{Coleman:1973jx}.  These logarithmic corrections can be large. Their size depends on the masses present in the theory,
which depend on the background field value, $\bar{h}$. In HI, the SM parameters are run up from the scale $\sim \bar{h}_{ew}$, where they are measured in the EW vacuum,
to the scale $\sim M_p/\sqrt{\xi}$, where inflation occurs.
This minimizes these large logarithmic corrections. The running is accomplished using the SM RG equations, which are  defined for running the Lagrangian parameters in energy.
The choice of a background field dependent renormalization scale $\mu^2 = \kappa(\bar{h}^2)$, used to minimize these logarithmic corrections, relates the running in energy to
running in the background field value.  The trajectory that the theory takes in $(\bar{h},E)$ space (were $E$ is the energy of the fluctuations of modes expanded around the background field value) depends on the choice of $\kappa(\bar{h}^2)$.

The discussion in Section \ref{nonlinearsection} makes clear that the interactions of the theory, and thus the RG equations,
depend in a nontrivial manner on the background field.\footnote{It is interesting to note that this is always the case,
and standard RG analyses that are running in energy alone implicitly assume that the background field is constant.}
In Ref. \cite{Salopek:1988qh} it is argued that by introducing the factor $s$
into the commutation relations of $h$ as
\bea
\left[h({\bf x}), \dot{h}({\bf y}) \right] = i s \, \hbar \, \delta^3({\bf x} - {\bf y}), \quad \quad \quad \! s =  \frac{1 + \xi \frac{\bar{h}^2}{M_p^2}}{1 + (1 + 6 \xi)\frac{\xi\bar{h}^2}{M_p^2}}, \eea
this effect can be incorporated. The form of $s$ is dictated by the kinetic mixing term, and the field redefinition to
take the theory to its canonical form. This factor is $\sim 1$ for $\bar{h} \ll M_p/\xi$ and the usual commutation relations are present. For
$M_p/\xi \leq \bar{h} \lesssim M_p/\sqrt{\xi}$, $s$ suppresses quantum loops involving $h$ by powers of $\sim 1/\xi$. The dependence on the background field, when $s$ is used to modify the SM RG's, includes corrections of order
$\mathcal{O}(\xi \, \bar{h}^2/ 16 \, \pi^2 \, M_p^2)$

Formally, the SM RG equations should be modified to include the background field dependence. This background field dependence is approximated in HI studies by
using two separate sets of RG equations. Below the scale $\Lambda_{ew}$, the SM running with the addition of a non-minimal coupling term
is used. Above the scale $\Lambda_{ew}$, the non-linear chiral lagrangian with a decoupled scalar singlet is used. This is a reasonable (although inexact)
method to approximate the background field dependence. We use this method in our numerical analysis in Section \ref{CMB}.\footnote{For some other recent numerical approaches see Ref. \cite{Allison:2013uaa}}.

Recently Ref. \cite{Jenkins:2013zja} calculated, the running of the SM parameters in the presence of higher dimensional operators, and noted that 
the running of the SM parameters themselves are modified by a background field dependent term.
To date, this fact been neglected in studies of HI. This difference is quadratically dependent on the background field value, and appears at one loop. Schematically the corrections are of the form
\bea\label{threshold}
\mu \, \frac{d c_4}{d \, \mu} = \frac{\lambda \, \bar{h}^2}{\Lambda^2} \, \frac{1}{16 \, \pi^2} \sum_i c_6^i.
\eea
Here $c_4$ stands in for a parameter in $\mathcal{L}_{SM}$, while
the sum over $i$ represents the sum over a subset of the dimension six operators, characterizing the degrees of freedom
integrated out. $\bar{h}$ is a parameter, not a field in this equation. See the Appendix where the exact results for 
Eqn. \ref{threshold} of Ref. \cite{Jenkins:2013zja} are reproduced for completeness.
These corrections scale as the ratio of the dimensionfull parameters
in the SM EFT, $m^2_h(\bar{h})/\Lambda^2$, where $m^2_h(\bar{h}) = 2 \, \lambda \bar{h}^2$.

When running the theory in background field space,  these corrections should be included.
Note that this modifies the running of the SM parameters below the scale present in unitarity violation arguments, which we take
as proximate to the scale $\Lambda$. Interestingly, around the scale $\bar{h} \sim \Lambda_{ew}$ these corrections dominate over the
background field dependence incorporated in HI analyses to date, so long as
\bea
 \lambda(\Lambda_{ew}) \gg \frac{1}{\xi(\Lambda_{ew})}.
\eea
The values of $\Lambda$ and $\xi$ at the scale of inflation are related through the WMAP normalization condition,
which gives
\bea
\frac{\lambda(\bar{h}_{inf}) \, M_p^4}{4 \, \xi^2(\bar{h}_{inf}) \, \epsilon(\bar{h}_{inf})} \simeq \left(0.0274 \, M_p \right)^4, \quad \quad \xi(\bar{h}_{inf}) \simeq 47000 \sqrt{\lambda(\bar{h}_{inf})}.
\eea
These corrections should be included if the $\bar{h}$ dependence of the RG equations is being approximated as in Ref. \cite{Bezrukov:2009db,Bezrukov:2013fka}.
This is another manner in which the scale $\Lambda_{ew}$ introduces UV sensitivity into the HI scenario.

It is easy to understand where these modifications of the SM RG equations originate. For example, loop diagrams with an internal Higgs field lead to a modification of the gauge field propagators. One finds \cite{Jenkins:2013zja}
a modification of the strong coupling running
\bea
\mu \frac{\rd g_3}{\rd \mu}  =  -\frac{g_3  \, m_H^2 }{4 \, \pi^2 \Lambda^2}  \, C_{H G}
\eea
due to the operator $Q_{HG} = H^\dagger \, H \, G_{\mu \, \nu} \, G^{\mu \, \nu}$.\footnote{Here we have modified the notation of
Ref. \cite{Jenkins:2013zja} to extract the factor of $1/\Lambda^2$} Corrections of this form are
also generated in an indirect manner, in re-normalizing the SM EFT, when the classical Higgs field EOM
\bea
D^2 H_k =  \frac{\lambda \, v^2}{2} H_k - 2 \lambda (H^\dagger H) H_k  -\overline q^j\, Y_u^\dagger\, u \, \epsilon_{jk} - \overline d\, Y_d\, q_k - \overline e\, Y_e\,  l_k.
\eea
is used to map obtained divergences to the retained EOM reduced operator basis. Here $j,k$ are $\rm SU(2)_L$ indices, and the remaining notation is
consistent with Ref. \cite{Jenkins:2013zja}. When we take the classical EOM for the Higgs field $H$, generalized to be the fluctuation around the classical background expectation
value $\langle H^\dagger \, H \rangle = \bar{h}^2$,  the sign of the leading term is flipped in the EOM above. An example of a term that receives such corrections is the running of $\lambda$, which receives one-loop contributions
to its running from sixteen higher dimensional operators, see Ref. \cite{Jenkins:2013zja}. Not all of these operators are
pure Higgs field operators. If one grants the assumption that some unknown mechanism controls the Higgs potential,
as in HI, there are still unknown corrections of this form that modify the running of the SM parameters, and introduce
UV sensitivity.

As the SM parameters run from the scale $\sim v$ to the scale $M_p/\sqrt{\xi}$, the relative size of the neglected corrections compared to
the SM one loop RG terms varies. Using the cut off scale determined in Ref. \cite{Burgess:2009ea}, in the low field regime
$\bar{h} \ll M_p/\xi$, this correction scales as
\bea
 \frac{\xi^2 \bar{h^2}}{M_p^2} \leq 1,
\eea
and is largest as $ \bar{h} \rightarrow M_p/\xi$. In fact at this scale, the power counting of the theory fails, in that,  the higher order terms of the form
$(\xi^2 \bar{h^2}/M_p^2)^n$ that also modify the running of the SM parameters, are no longer suppressed. This indicates the clear UV sensitivity that this scale introduces.
In the intermediate field region $M_p/\xi \ll \bar{h} \ll M_p/\sqrt{\xi}$, using the cut off scale determined in Ref. \cite{Bezrukov:2010jz}, this correction scales as
\bea
\frac{m_H^2(\bar{h}^2) \sum_i c_i}{\Lambda^2} \sim \frac{\sum_i c_i/g^2_{\star}}{16 \, \pi^2} \, \frac{\lambda \, M_p^2}{\xi^2 \, 3} \, \frac{1}{\bar{\chi}^2},
\eea
as the scale of unitarity violation is expected to be  $M_\star \sim 4 \, \pi \, g_\star \, \bar{h}$  with $g_{\star} < 1$, in this region.
Here $g_\star$ is a general parameter that is determined by the exact spectrum and dynamics of the UV theory. In particular, the lightest state integrated out
that contributes to a particular operator can determine $g_{\star} $ in some scenarios. Note that these RG corrections are suppressed in the chiral phase
at large $\xi$. In the numerics presented in Section \ref{nonlinear} we will neglect this further UV sensitivity.

The systematic renormalization results of Refs.\cite{Jenkins:2013zja,Jenkins:2013wua,Alonso:2013hga} are calculated for the SM with a linear realization of $\rm SU_L(2) \times U_Y(1)$, and  performed in flat space, where $\bar{h}_{ew} = v$. 
Here we have taken the classical EOM for the Higgs field $H$, generalized to be the fluctuation around the classical background expectation
value $\bar{h}$.  There are further corrections to the renormalization of the SM EFT, due to the coupling of the theory to gravity, and when renormalizing the theory in curved space. Further, the EOM are also modified, with the non-minimal coupling leading to extra terms\footnote{See Ref. \cite{George:2013iia} for a discussion of these terms in the context of singlet scalar non-minimally coupled to gravity.} $ \propto \dot{\mathcal{H}} + 3 \mathcal{H}^2$. As our purpose is just to show the explicit UV sensitivity introduced in the RG evolution by the effects we retain,
we neglect these further modifications.\footnote{Note that we have also neglected corrections of the form considered in this section to the running of $\xi$. The
complete renormalization of the SM EFT in curved space is beyond the scope of this work. This
is potentially of interest as the running of $\xi$ can be related to the running of $\lambda$ due to the requirement that the effective potential at its extremum
being renormalization scale independent. The numerical sensitivity to higher dimensional operators in HI  is still present even if the effect of the higher dimensional operators on the running of $\xi$ is assumed to cancel the running of $\lambda$. We have explicitly checked this is the case.}

\section{Perturbations, Linear and Nonlinear}\label{CMB}

One of the challenges to HI, is the measured Higgs mass. Taking the central value of the Higgs mass, and the central value for $m_t$ and $\alpha_s$,
the parameter $\lambda$ runs negative far before the scale at which inflation occurs.

A shift in the SM parameters at either the EW scale or at intermediate scales can allow HI to occur, as illustrated in Fig 2.
\begin{figure}[t]
  \hfill
  \begin{minipage}[t]{.49\textwidth}
    \begin{center}
      \epsfig{file=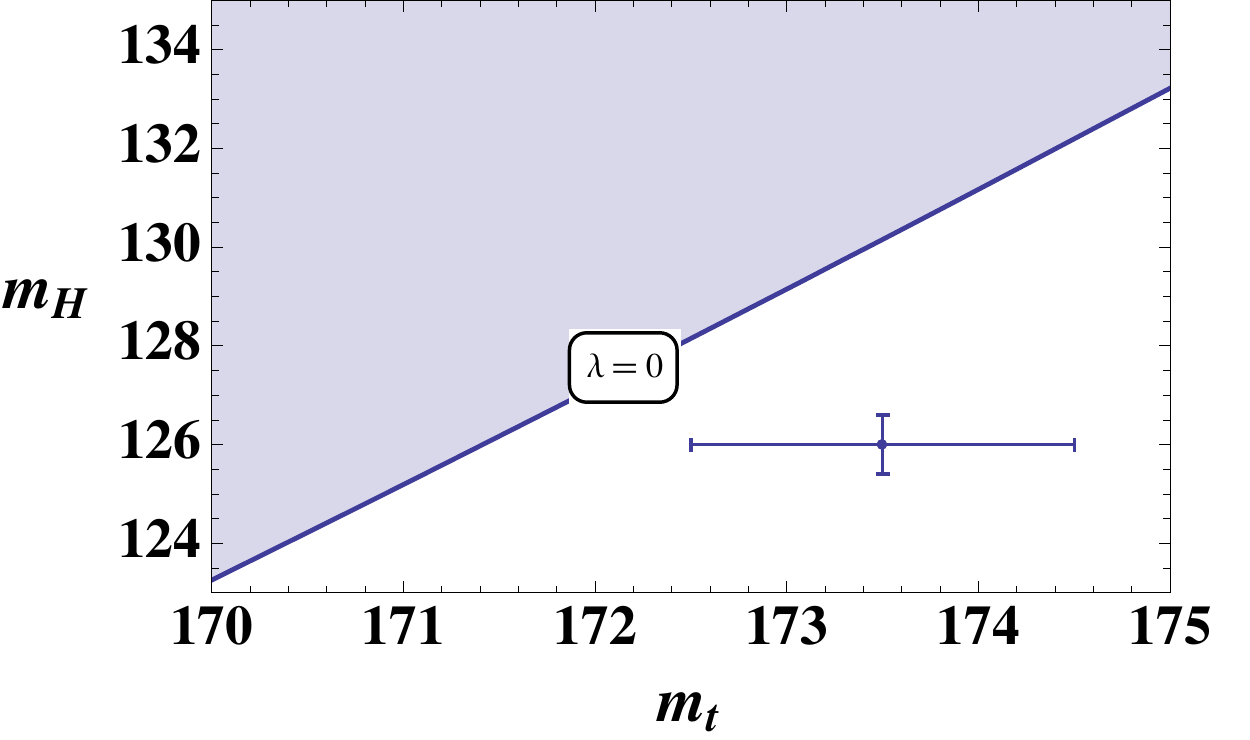, height=2.2in, width=3in}
      \label{enhanced}
    \end{center}
  \end{minipage}
  \hfill
  \begin{minipage}[t]{.49\textwidth}
    \begin{center}
      \epsfig{file=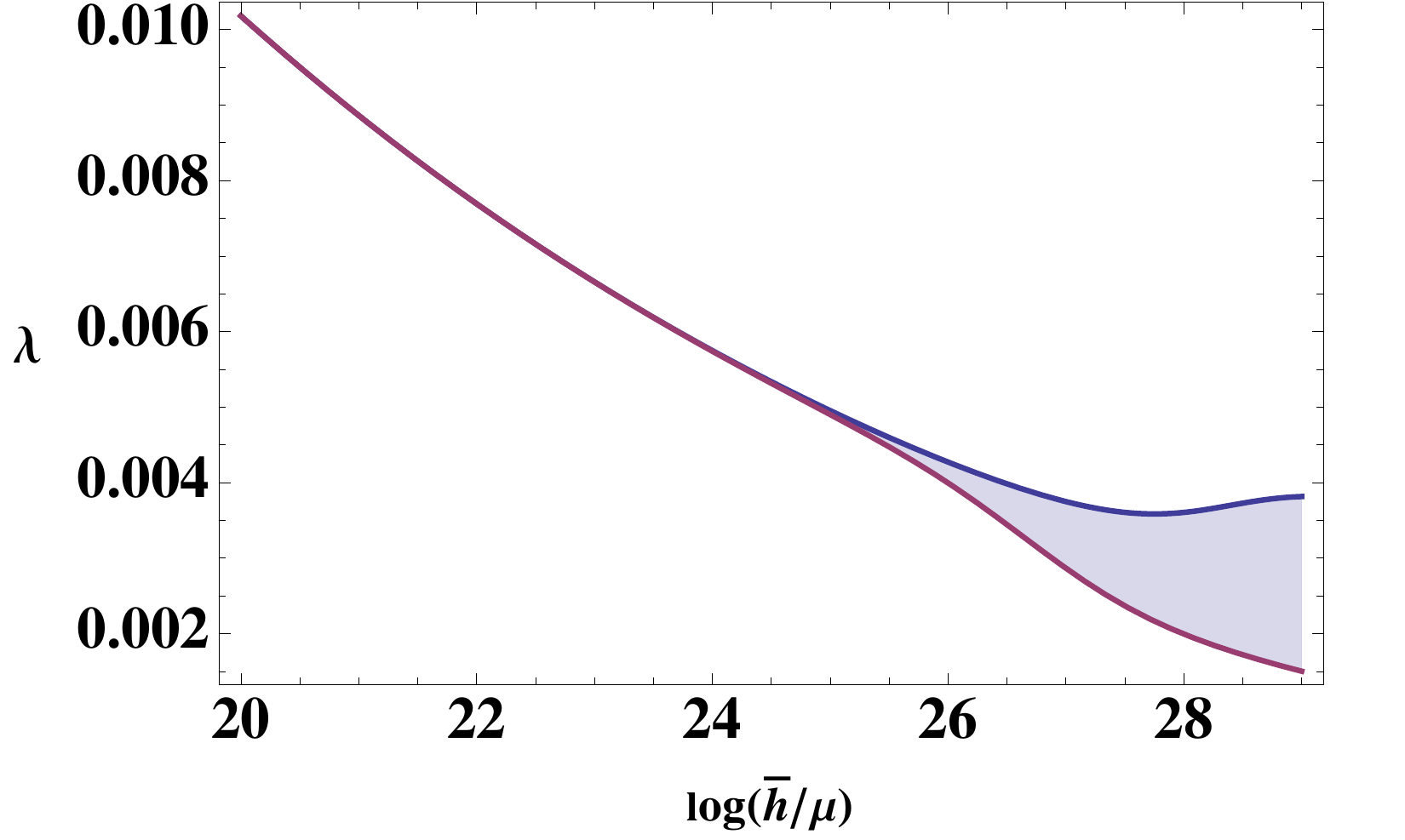, height=2.15in, width=3in}
      \label{step}
    \end{center}
  \end{minipage}
  \hfill
   \caption{\small{(Left) The initial conditions that separate $\lambda >0$ from $\lambda < 0$ at the scale $M_{pl}/\sqrt{\xi}$, taken as a proxy for whether or not Higgs inflation can connect to the EW vacuum once it ends. Above the line $\lambda >0$. Also shown is the one sigma error bar range for the top quark mass and the Higgs mass.
   For the later we use the number reported in Ref. \cite{Aad:2012tfa}, for the former we use the PDG number. (Right) The spread in the values for the quartic coupling induced by the RG corrections given $m_t = 170.95\, {\rm GeV}, m_H = 125.66\, {\rm GeV}$.}}
\end{figure}
We have checked that the effect discussed in this paper, the modification of the running of the SM parameters due to dimension six operators, does not significantly expand the range of allowed Higgs masses that allow sucessful inflation,
assuming the top quark mass takes on its central value shown in Fig.2. The shift in the allowed Higgs for $\lambda > 0$ at the scale of inflation is $\lesssim 1 \, {\rm GeV}$.\footnote{Note added: Subsequent to this work the authors of \cite{Bezrukov:2014bra} use similar techniques to explore the possibility that similar threshold corrections at $m_p/\xi$ from dimension six operators could be used to reconcile the stability of the Higgs potential in the inflationary regime with top (pole) masses closer to the central value.}

In the following Sections,  we first consider small linear perturbations to understand how the CMB parameters scale with changes in the effective parameters at $\sim M_p/\sqrt{\xi}$. We then consider the full non-lineary perturbed renormalization group running to illustrate the UV sensitivity with numerical results. Due to the non-linear nature of the RGE evolution (and the secular growth of small perturbations in the parameters from running over many orders of magnitude), the later approach is necessary. The linear perturbation results are only presented to offer some limited analytic intuition on the UV sensitivity.

\subsection{Linear Perturbations}

Assume that their exists a set of parameters $\xi, \lambda$ that allow inflation to occur, and $\epsilon, \eta$ the parameters that characterize the resulting slow roll phase:
\bea
\label{srp1}
\epsilon &=& \frac{M_p^2}{2}\left(\frac{U_{,\chi}}{U}\right)^2 = \frac{M_p^2}{2}\left(\frac{U'}{U}\right)^2\frac{1}{\chi'^2}\\ \label{srp2}
\eta &=& M_p^2\frac{U_{,\chi\chi}}{U} = M_p^2\frac{U''}{U}\frac{1}{\chi'^2} - M_p^2\frac{U'}{U}\frac{\chi''}{\chi'^3} 
\eea
where these parameters are defined with respect to the canonically normalized field, which we express in terms of the singlet $\bar h$ through the change of variable $\chi' = d\chi/d\bar h$, primes denoting derivatives w.r.t. $\bar h$. 

Label the parameters that correspond to successful inflation as  $\epsilon_0, \eta_0$. Now consider a perturbation of these parameters in the semi-classical analysis. Assume the changes in the CMB parameters can be approximated by a linear
perturbation, neglecting higher order terms, then
\bea
\delta n_s =  - 6 \,  \frac{\delta \epsilon}{\epsilon_0} + 2 \,  \frac{\delta \eta}{\eta_0}, \quad  \quad \quad \quad  \quad \quad  \quad \delta r = 16 \, \frac{\delta \epsilon}{\epsilon_0}.
\eea
We will restrict ourselves to the case where $\xi \gg 1$. This allows some simplification of the resulting equations.
Let $\mu^2 = \kappa(\bar{h}^2)$, but  the specific choice of $\kappa(\bar{h}^2)$ will be left unfixed. Two possible choices are \cite{Bezrukov:2013fka}
\bea\label{fphichoice}
\kappa(\bar{h}^2) = \frac{y_t^2}{2} \bar{h}^2, \quad \quad \kappa(\bar{h}^2) = \frac{y_t^2 \, \bar{h}^2
}{2 \left(1 + \xi \bar{h}^2/M_p^2 \right)}.
\eea
Which correspond to minimizing the logarithms in Eqn \ref{CWcorr} due to the top quark mass, in the Jordan or Einstein frames.
The potential in the Einstein frame, with the scale $\mu$ chosen so that corrections to the CW potential are suppressed, is given by
\bea
V_E(\chi) = V_0 \left[1 - e^{- \beta \chi} \right]^2 + \cdots, \quad \quad V_0 =  \frac{\lambda(\kappa({\chi}^2)) \, M_p^4}{4 \, \xi^2(\kappa({\chi}^2))}.
\eea
The field derivative of $\mu$ in the large $\xi$ limit, for large field values during inflation is given by
\bea
\frac{d \, \log \mu/M_p}{d \hat{\bar{\chi}}} 
%&=& \frac{1}{2} \, \frac{f'(\bar{\phi})}{f(\bar{\phi})} \frac{d \bar{\phi}}{d \chi},  \\
&\simeq& \frac{\kappa'(\hat{\bar{\chi}})}{\kappa(\hat{\bar{\chi}})}  \frac{M_{p} \, \beta}{4 \, \sqrt{\xi}} \left[ C_\chi^{1/2} +  C_\chi^{-1/2} \right] 
\eea
where $C_\chi =  -1 + e^{M_p \,\beta \hat{\bar{\chi}}}$. Here $\hat{\bar{\chi}}$ is the vev of the field $\chi$ normalized to $M_p$.
The slow roll parameters are given by
\bea
\frac{\epsilon}{M_p^2 \, \beta^2} &\simeq& \frac{2}{C_\chi^2} + \left[\frac{1}{V_0} \frac{d V_0}{d \log \mu/M_p}\right] \frac{\kappa'}{2 \, \kappa \, \sqrt{\xi}}  \, (C_\chi^{-1/2}+ C_\chi^{-3/2})+ \frac{1}{32 \, V_0^2} \left[\frac{d V_0}{d \log \mu}\right]^2 \left(\frac{\kappa'}{\kappa \, \sqrt{\xi}}\right)^2 \, \left[ C_\chi^{1/2} +  C_\chi^{-1/2} \right]^2 , \nonumber \\
%%%%%%%%%%%%%%%%%%%%%%%%%%%%%%%%%%%%%%%%
\frac{\eta}{M_p^2 \,\beta^2} &\simeq& \frac{2 (C_\chi -1)}{C_\chi^2} +  \left[\frac{1}{V_0} \, \frac{d^2 V_0}{d \log^2 \mu/M_p} \right] \, \left[\frac{\kappa'}{4 \, \kappa \, \sqrt{\epsilon}} \right]^2 
\left[ C_\chi^{1/2} +  C_\chi^{-1/2} \right]^2, \nn 
&-&  \left[\frac{1}{V_0}  \, \frac{d V_0}{d \log \mu} \right] \, \frac{1}{M_p \, \beta} \left[C_\chi^{1/2} +  C_\chi^{-1/2} \right] \left[\sqrt{\xi} \, \left[\frac{\kappa'}{2 \, \kappa \, \sqrt{\xi}} \right]^2  - \frac{\kappa''}{4 \, \kappa \,  \sqrt{\xi}}\right] \\
&+& \left[\frac{1}{V_0} \, \frac{d V_0}{d \log \mu} \right] \, \left[\frac{\kappa'}{4 \, \kappa \,  \sqrt{\xi}} \right] \, \frac{\left[C_\chi^2 + 4 \, C_\chi + 3 \right]}{C_\chi^{3/2}}. \nonumber
\eea
These expressions can be simplified somewhat. Take the large $\xi$ limit, assuming the scaling
\bea
\frac{\kappa'}{\kappa} \sim \sqrt{\xi}, \quad \quad \frac{\kappa''}{\kappa} \sim \xi, 
\eea
which is consistent with the choices for $f$ in Eqn \ref{fphichoice}. Further, in perturbation theory
\bea
\frac{d^2 V_0}{d \log^2 \mu}  \ll \frac{d V_0}{d \log \mu}, \quad \quad \left(\frac{d V_0}{d \log \mu}\right)^2  \ll \frac{d V_0}{d \log \mu}
\eea
so that the leading corrections are given by
\bea
\frac{\epsilon}{M_p^2\, \beta^2} &\simeq&\frac{2}{C_\chi^2} + \left[\frac{M_p}{V_0} \frac{d V_0}{d \log \mu}\right] \frac{\kappa'}{2 \, \kappa \, \sqrt{\xi}}  \, (C_\chi^{-1/2}+ C_\chi^{-3/2}), \\
\frac{\eta}{M_p^2\, \beta^2} &\simeq&\frac{2 (C_\chi -1)}{C_\chi^2}  - \left[\frac{1}{V_0}  \, \frac{d V_0}{d \log \mu} \right] \, \frac{\left[C_\chi^{1/2} +  C_\chi^{-1/2} \right]}{M_p \, \beta} \left[\sqrt{\xi} \, \left[\frac{\kappa'}{2 \, \kappa \, \sqrt{\xi}} \right]^2  - \frac{\kappa''}{4 \, \kappa \,  \sqrt{\xi}}\right]
\eea
Also we note that
\bea
\frac{1}{V_0} \,  \frac{d V_0}{d \log \mu} = \frac{\beta_\lambda}{\lambda}  - 2  \, \frac{\beta_\xi}{\xi}. 
%\quad \quad \quad
%\frac{1}{V^2_0} \,  \left[\frac{d V_0}{d \log \mu}\right]^2 = \frac{\beta_\lambda^2}{\lambda^2}  - 4  \, \frac{\beta_\xi}{\xi} \, \frac{\beta_\lambda}{\lambda} + 4  \, \frac{\beta^2_\xi}{\xi^2}.
\eea

The effect of the RG corrections that we include is to introduce extra terms in the $\beta$ functions. The change in the running
of $\xi$ can be (mostly) absorbed into this parameters normalization. This simple analysis indicates that $\delta \epsilon/\epsilon_0 \sim \delta \eta/\eta_0$.
In the detailed numerics presented in the next Section, we find this is the case.  Due to the fact that $\eta_0 \gg \epsilon_0$, for the plots shown, the
smearing out of the prediction is mostly for $n_s$ while leaving $r$ essentially unchanged. These results also indicate that the effect should be quite small, 
where the simple linear perturbation theory  considered here is not breaking down.

\subsection{Renormalization group running}\label{nonlinear}
In what follows, we implement the prescription laid out in Ref. \cite{Bezrukov:2009db} to compute the renormalization group improved potential during inflation. In order to do so, we must first run the standard model parameters up to the scale $M_{p}/\xi$ to two loop order, with initial couplings defined at top pole mass, whose values at NNLO have recently been computed in Ref. \cite{Buttazzo:2013uya} in the $\rm \overline{MS}$ scheme\footnote{Here $\{\lambda, y_t, g_1, g_2, g_3\}$ are the quartic self-coupling of the Higgs, the top quark yukawa and the $\rm SU(2)$, $\rm U(1)$ and $\rm SU(3)$ gauge couplings, respectively. The value of $\alpha_3(m_Z)$ is held fixed at $0.1184$ as is the pole mass of the $W$ boson.}: 
\bea
\nonumber y_t(\mu = m_t) &=& 0.93558 + 0.00550\left(\frac{m_t}{\rm GeV} - 173.1\right) \pm 0.00050_{th} \nonumber\\
g_1(\mu = m_t) &=& 0.35761 + 0.00011\left(\frac{m_t}{ \rm GeV} - 173.1\right) \nonumber\\
g_2(\mu = m_t) &=& 0.64822 + 0.00004\left(\frac{m_t}{\rm  GeV} - 173.1\right)\\
\nonumber g_3(\mu = m_t) &=& 1.1666 - 0.00046\left(\frac{m_t}{\rm GeV} - 173.1\right)\\
\lambda(\mu = m_t) &=& 0.12711 + 0.00206\left(\frac{m_h}{\rm GeV} - 125.66\right) \nonumber - 0.00004\left(\frac{m_t}{\rm GeV} - 173.1\right) \pm 0.00030_{th} \nonumber
\eea  
As discussed in the previous Sections, at the scale $M_p/\xi$, the singlet component of the Higgs starts to effectively decouple from all other fields, leaving us with the non-linearly realized chiral EW theory plus the singlet scalar, our inflaton. We compute this field's effective CW potential (also evaluated at top pole mass so as to minimize the logarithms) at the scale of inflation. We follow Ref.\cite{Bezrukov:2009db} and use the one-loop expression for the CW potential.
We run the couplings of the tree level part of the potential at one loop up to the scale of inflation, with the modified beta functions of the chiral EW theory\footnote{Where the running of the couplings relative to the SM case differs due to the absence of any off-shell Higgs propagators in the loops. We refer to Ref. \cite{Bezrukov:2009db} for the one-loop beta functions in the chiral phase.}. The result will be the Einstein frame RG improved effective potential
\be
V_{E}(\bar\phi) = \frac{\lambda(\mu(\bar h)) \bar h^4}{\left(1 + \frac{\xi(\mu(\bar h))\bar h^2}{M_p^2}\right)^2} + \cdots
\ee
where through either choice in Eqn. (\ref{fphichoice}) for the renormalization scale $\mu$ itself depends on $\bar h$. From this, deriving CMB observables uses Eqns. (\ref{srp1}), (\ref{srp2}) and (\ref{fphichoice}).  (We choose the renormalization scale consistent with perscription one in Ref.\cite{Bezrukov:2009db}, which corresponds to the Right hand Equation in Eqn \ref{fphichoice}.) Inflation is taken to end when $\epsilon = 1$ and all CMB observables are to be evaluated at the time at which the COBE normalization scale $k = 0.002 Mpc^{-1}$ exits the horizon, some $N_e$ e-folds before the end of inflation, where
\be
N_e = \frac{1}{\sqrt 2 M_p}\int^{\bar h_f}_{\bar h_i} \frac{\chi'}{\sqrt\epsilon} \, d\bar h
\ee   
The only difference in our implementation is that we now include the corrections to the RG running in the standard model phase of the theory, schematically denoted as
\bea
\nonumber
\mu \frac{d g_i}{d \, \mu} &:=& \Delta \beta_{g_i} = -  \frac{\lambda \, \bar{h}^2}{2 \, \pi^2 \, \Lambda^2} \, g_i \, C_{(i)}\,, \\
\mu \frac{\rd}{\rd \mu} \lambda &:=& \Delta \beta_{\lambda} = \frac{\lambda \, \bar{h}^2}{16 \pi^2 \Lambda^2}  \, \sum_j \, C_{(j)}\,, \\
\mu \frac{\rd}{\rd \mu} y_t &:=& \Delta \beta_{y_t} = \frac{\lambda \, \bar{h}^2 }{16\pi^2 \, \Lambda^2}\, y_t \, \sum_k \,C_{(k)} \nonumber
\eea
where $i$ runs over $1\leq i \leq 3$, while  and $C_{(j,k)}$ are a sum over other couplings and their respective Wilson coefficients (see Appendix A). We
include a profile function for the cut-off $\Lambda$ that depends on $\bar h$ 
\bea
\label{lamdep}
\Lambda^2(\bar h) = \frac{(M_{p}^2 +  \xi \bar{h^2} + 6 \xi^2  \bar{h^2})^2}{\xi^2 (M_{p}^2 + \xi \bar{h^2})},
\eea
This is consistent with the cut off scales discussed in the previous sections, and Eqn. \ref{two} \cite{Bezrukov:2010jz}. 
Note that the cut off scale quoted above, obtained in the Jordan frame, is consistent with an asymptotic constant value in terms of planck mass units as described in Section \ref{cutoffs}.
We scan over various values of $C_{(i,j,k)}$ consistent with variations of the constituent Wilson coefficients ranging over values of order unity, where for example $C_j$ being an aggregate of several independent co-efficients (\ref{Adef}), we scan over a range that is the root mean square of the individual variations.   
The UV dependent terms in the RG  give the differential equations a "kick" just around the scale $M_p/\xi$, which effectively serves to smear out the initial conditions for the running of the couplings in the chiral phase, whose RGE's we patch to at $\bar h = M_p/\xi$ and run up to the scale of inflation. This spread in the possible initial values for the couplings at the commencement of the chiral phase represents the irreducible theoretical uncertainty associated with not knowing the UV completion of the SM non-minimally coupled to gravity, which then propagates into an uncertainty in our computation of cosmological observables. 

Fig. 3 shows how these corrections can effect the effective potential and the predictions for the spectral tilt and the scalar to tensor ratio. In each run over a particular set of Wilson coefficients, we set $N_e = 57.7$ and require that the effective potential thus computed be COBE normalized at $k = 0.0002 Mpc^{-1}$, tuning the initial value of $\xi$ accordingly. It is possible to visually identify that although COBE normalization partly nullifies the dependence of the spectral properties of the CMB on the value of potential during inflation (depending as it does only on $V_{inf}/\epsilon$), the precise shape of the potential is affected by the kicks in the RG running induced by the unknown UV dependent dimension six operators. The smearing of the running further towards the red (lower values of $n_s$) can be readily understood from the fact that the shape of the effective potential is typically made steeper, rather than shallower once one scans over the unknown Wilson coefficients. The tensor to scalar ratio also ranges over $O(10^{-3})$ to $O(10^{-4})$ as you scan over the Wilson coefficients, though at the scale of the plot this is essentially degenerate with the axis.    
\begin{figure}[t]
  \hfill
  \begin{minipage}[t]{.49\textwidth}
    \begin{center}
      \epsfig{file=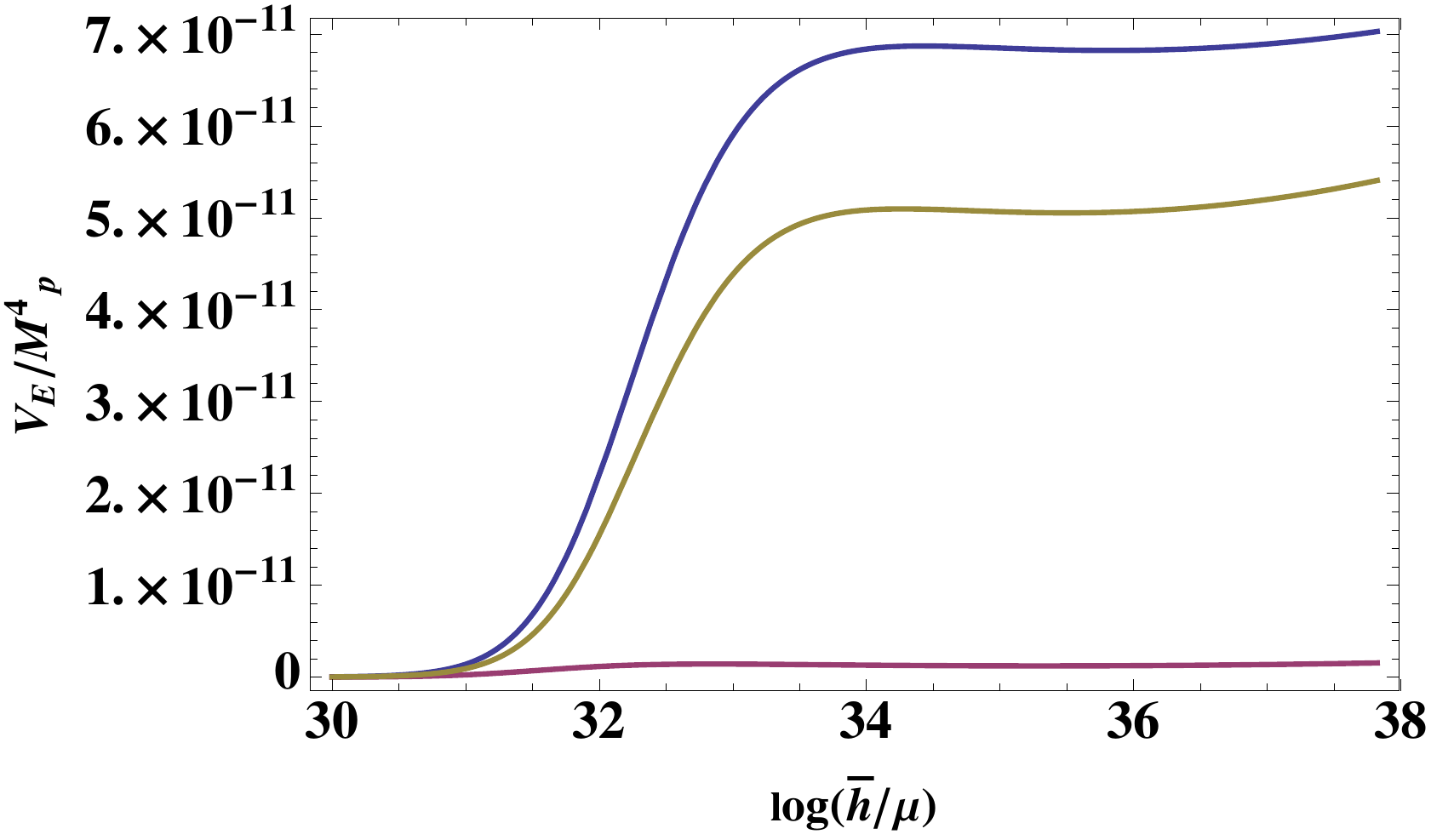, height=2.25in, width=3.0in}
      \label{enhanced}
    \end{center}
  \end{minipage}
  \hfill
  \begin{minipage}[t]{.49\textwidth}
    \begin{center}
      \epsfig{file=plot1.pdf, height=2.2in, width=3.2in}
      \label{step}
    \end{center}
  \end{minipage}
  \hfill
   \caption{\small{RG improved potential and spectral index vs. r for $m_t = 170.95\, {\rm GeV}, m_H = 125.66\, {\rm GeV}$. On the right plot $r$ ranges from $r = 1 \times 10^{-3}$ to $3 \times 10^{-5}$ as the spectral index changes from $n_s = 0.957$ to $0.885$, essentially indistinguishable from the x-axis. The red dot represents the prediction with no corrections terms in the RG equations due to Higher D operators, with $n_s = 0.955$. On the left plot, the effective potential is plotted for the two outliers of the scan over Wilson co-efficients, along with the RG improved potential for the case of no higher D operator effects in the RG equations.}}
\end{figure}

\section{Conclusions}\label{conclusions}

We have re-examined the issues of UV sensitivity in inflationary single field models, focusing on the interesting case of Higgs Inflation. The effect of unknown higher dimensional operators were shown to have an observable impact on CMB predictions in this case. This is an irreducible theoretical uncertainty (in our view) until the exact UV that completes
the theory is specified. It is not sufficient to banish higher dimensional operators that are composed only of Higgs fields in models of this form to maintain predictivity. The higher dimensional interactions of the same dimension extensively mix, at sufficient loop order. Further the higher dimensional operators mix down and modify the SM parameter running in a manner that depends on the background field value. This introduces UV sensitivity at the scale $M_{p}/\xi$ through the RG equations, in an interesting manner. The requirement of an exponentially flat potential makes some inflationary models particularly sensitive to these effects.

Note added on recent developments: The recent BICEP2 measurement \cite{Ade:2014xna} of a comparatively large primordial tensor fluctuation, $r = 0.20^{+ 0.07}_{-0.05}$, puts some pressure on the Higgs Inflationary scenario which predicts smaller $r$ for inflation driven by the exponential rollout from the asymptotically constant Einstein-frame potential at large fields. (See Ref. \cite{Cook:2014dga}, for example, for a recent discussion). Other recent works \cite{Hamada:2014iga, Bezrukov:2014bra} counter this with ways to evade the problem in special parts of parameter space. (For instance one can choose special values for $m_t$ and $m_h$ --- though not within the one-sigma measured values --- such that the critical point in the SM Higgs potential occurs at scales similar to those required by the BICEP2 measurements.) Once this is done a larger value of $r$ can be obtained, potentially consistent with the BICEP2 results.

While we are willing to take these claims at face value, we would make the following comment: the larger value of $r$ so obtained comes at the expense of a much smaller value of $\xi$: $\xi \sim 10$ rather than $\sim 10^4$. This is worrisome for the control of approximations used, the point initially raised in Ref.~\cite{Burgess:2009ea}, since it is precisely the large value of $\xi$ that provides the hierarchy between the Planck scale $M_p$, the large-field unitarity scale, $\Lambda \sim M_p/\sqrt\xi$, and the inflationary Hubble scale $H \sim M_p/\xi$. For $\xi$ of order 10 the unitarity scale is only 3 times larger than the Hubble scale during inflation, and both are uncomfortably close to the Planck scale. The effects of higher-dimension operators emphasized in this article are also a concern in this case; with the range of predicted values for both $r$ and $n_s$ being much larger than their measured errors. In general, smaller $\xi$ implies close proximity to the UV `Planck wall', thereby sharpening all issues associated with the unknown UV completion at these scales.

\section*{Acknowledgements}

We thank Ignatios Antoniadis, Jan Hamann, Hyun-Min Lee, Andrea de Simone, Sergey Sibiryakov and Witold Skiba for useful discussions and correspondence. SP is supported by a Marie Curie Intra-European Fellowship of the European Community's 7'th Framework Program under contract number PIEF-GA-2011-302817. CB's research is supported in part by funds from the Natural Sciences and Engineering Research Council (NSERC) of Canada and from Perimeter Institute for Theoretical Physics. Research at the Perimeter Institute is supported in part by the Government of Canada through Industry Canada, and by the Province of Ontario through the Ministry of Research and Information (MRI). We also thank Marieki Postma and Jacopo Fumagalli for pointing out an error in the linear perturbation section. However we note that the linear perturbation section does not effect the remainder of the paper, and was simply for building some minor intuition.

\appendix
\section{Higgs-axion and Higgs-graviton mixing}
\label{HiggsAx}

A toy model for Higgs-graviton mixing is the case of Higgs-axion kinetic mixing, with Lagrangian density
\bea
 \cL &=& - \frac12 \, ( \partial h)^2 - \frac12 \, m^2 \, h^2 - \frac12 \, (\partial a)^2 - \frac{v}{f} \, h \Box a  \nn\\
 &=& - \frac12 \, ( \partial h)^2 - \frac12 \, m^2 \, h^2 - \frac12 \, (\partial a)^2 + \frac{v}{f} \, (\partial_\mu h )(\partial^\mu a) \,.
\eea
Here the axion's shift symmetry, $a \to a + f$, keeps it massless (much like general coordinate invariance keeps the graviton massless). In the Higgs-inflation story $f$ is the analogue of $M_p/\xi$, since $4\pi f$ would be the unitarity scale for the axion alone..

This is diagonalized by taking
\be
 a = \psi + \frac{h v}{f} \,,
\ee
so that
\be
 \cL = - \frac12 \, \left( 1 - \frac{v^2}{f^2} \right) (\partial h)^2 - \frac12 \, (\partial \psi)^2 - \frac12 \, m^2 \, h^2 \,,
\ee
and so, canonically normalizing gives $h = \chi/\sqrt{1 - v^2/f^2}$ gives
\be
 \cL = - \frac12 \, (\partial \chi)^2 - \frac12 \, (\partial \psi)^2 - \frac{m^2}{2(1 - v^2/f^2)} \, \chi^2 \,.
\ee
This shows the physical Higgs mass gets increased to
\be
 m^2_h = \frac{m^2}{1 - v^2/f^2} \,,
\ee
and all $h$ couplings with SM matter similarly get increased. {\em e.g.}
\be
 - \frac12 \, g^2 (v + h)^2 \, W^*_\mu W^\mu = - M_\ssW^2 \left( 1 + \frac{\chi}{\sqrt{1 - v^2/f^2}} \right)^2 \, W^*_\mu W^\mu \,.
\ee
Notice one would never be tempted to entertain the regime $v > f$ in this model. 

Graviton-Higgs mixing is very similar, but with two important changes. First, the metric trace, $h$, has {\em negative} kinetic term, $\cL \sim - \frac12 \, h \Box h$, and this turns the factors of $1 - v^2/f^2$ into $1 + v^2/f^2$, thereby suppressing the couplings and allowing us to believe the $v \gg f$ limit. Second, gauge invariance allows the nominally unstable mode, $h$, to be gauged away.

\section{Dimension six operator corrections}
In the basis of (non-redundant) operators defined in Ref. \cite{Grzadkowski:2010es}, the mixing of all dimension six effective operators, including non trivial flavour structure, into the running of dimension four operators has been calculated at one loop in Refs. \cite{Grojean:2013kd,Jenkins:2013zja,Jenkins:2013wua,Alonso:2013hga}. The beta functions that determine the running of the SM gauge couplings are modified as \cite{Jenkins:2013zja}
\bea
\nonumber
\mu \frac{d g_3}{d \, \mu} &=& \frac{\lambda \, \bar{h}^2}{2 \, \pi^2 \, \Lambda^2} \, g_3 \, C_{HG}, \\
\mu \frac{d g_2}{d \, \mu} &=& \frac{\lambda \, \bar{h}^2}{2 \, \pi^2 \, \Lambda^2} \, g_2 \, C_{HW}, \\
\nonumber
\mu \frac{d g_1}{d \, \mu} &=& \frac{\lambda \, \bar{h}^2}{2 \, \pi^2 \, \Lambda^2} \, g_1 \, C_{HB}.
\eea
The notation for the operators differs from Ref. \cite{Jenkins:2013zja} in that an explicit factor of $1/\Lambda^2$ has been
factored out of the Wilson coefficients $C_i$. Also the sign of the contribution has been flipped, as we expand around the large classical background field, not the EW vev. The corrections to the SM running of the quartic coupling and the Yukawa matrices are given by \cite{Jenkins:2013zja}
\bea
\mu \frac{\rd}{\rd \mu} \lambda &=& -\frac{\lambda \, \bar{\phi}^2}{16 \pi^2 \Lambda^2}  \biggl[ A_{\lambda} + B_{\lambda} + D_{\lambda} \biggr]\,, \\
%%%%%%%%%%%%%%%%%%%%%%%%%%%%%%%%%%%%%%%%%%%%
\mu \frac{\rd}{\rd \mu}  [Y_u]_{rs} &=&  - \frac{\lambda \, \bar{\phi}^2 }{16\pi^2 \, \Lambda^2}  \biggl[ A_{rs}^{y_u} + B_{rs}^{y_u}  \biggr]. \nonumber
\eea
The parameters $A_i,B_i,D_i$ depend on the UV completion and are given by a straightforward modification of the
results in Ref. \cite{Jenkins:2013zja}. Here the number of colours is $N_c=3$, $y_H = 1/2$ and $c_{F,3}=4/3$, $c_{F,2} = 3/4$ and $c_{A,2}=2$. 
The contributions that come from diagrams with no internal Higgs
fields in the loop are grouped into the $A_a$ coefficients, whereas those that contain one and two internal Higgs fields are grouped into the $B_i$ and the $D_i$ coefficients respectively. See Ref. \cite{Grzadkowski:2010es,Jenkins:2013zja} for more details on the operator basis used. The $A_i,B_i,D_i$  are given in terms of the unknown Wilson coefficients $C_i$ as  \cite{Jenkins:2013zja}
\bea\label{Adef}
A_{\lambda}  &=& - 3 \, g_2^2 \, C_{H D}   + 4 \, \eta_1 + 4 \eta_2 + 24 \, g_1 g_2 y_H C_{HWB}  - 6 \,C_{A,2} \, g_2^3 \, C_{W} +\frac83 g_2^2 C^{(3)}_{\substack{H l \\ tt}} +\frac83 g_2^2 N_c C^{(3)}_{\substack{H q \\ tt}},  \\ \nonumber
A_{rs}^{y_u}  &=&  -4 \left( C^{(1)*}_{\substack{qu \\ sptr}} + c_{F,3} C^{(8)*}_{\substack{qu \\ sptr}} \right) [Y_u]_{tp}
- 2 C^{(1)*}_{\substack{lequ \\ ptsr}} [Y_e^*]_{tp}+ 2 N_c C^{(1)*}_{\substack{quqd \\ srpt}} [Y_d]^*_{tp}
+ \left( C^{(1)*}_{\substack{quqd \\ prst}} + 2 \, c_{F,3} C^{(8)*}_{\substack{quqd \\ prst}}  \right) [Y_d]_{tp}^*,
\eea
\bea
\label{bdef}
B_{\lambda} &=& 24 C_H 
+ 24 \, \left(g_2^2 c_{F,2} C_{HW} + g_1^2 y_H^2 C_{HB} -\frac12 \, g_1 g_2 y_H C_{HWB}  + \frac14 \,C_{A,2} \, g_2^3 \, C_{W} \right) \\
&\,& - 8 \, \lambda C_{Hbox} + 4 \left(\lambda + 3 \, g_1^2  y_H^2 \right) C_{H D}, \nn 
B_{m_H} &=& - 16  \, C_{Hbox} + 8 \, C_{H D},\nonumber \\ 
B_{rs}^{y_u} &=& 6 \, C_{\substack{uH \\ sr}}^* -\left(2 C_{Hbox} - C_{H D} \right) [Y_u]_{rs} - 2 [Y_u]_{rt} \left( C^{(1)}_{\substack{H q \\ ts}} +3 C^{(3)}_{\substack{H q \\ ts}} \right)+ 2 C_{\substack{H u \\ rt}} [Y_u]_{ts}  - 2 C_{\substack{H ud \\ rt}}  [Y_d]_{ts}, \nonumber
\eea
\bea
\label{cdef}
D_{\lambda}  &=& - 56  \, C_{Hbox} + 20 \, C_{H D}, \\
\eta_1 &=& \left( \frac12 N_c C_{\substack{dH \\ rs}} [Y_d]_{sr} +  \frac12 N_c C_{\substack{uH \\ rs}} [Y_u]_{sr} + \frac12 C_{\substack{eH \\ rs}} [Y_e]_{sr} \right) + h.c.\,, \\
\eta_2 &=&  -2 N_c C_{\substack{H q \\ rs}}^{(3)} [Y_u^\dagger   Y_u ]_{sr} -2 N_c C_{\substack{H q \\ rs}}^{(3)} [Y_d^\dagger  Y_d  ] _{sr} + N_c C_{\substack{H ud \\ rs}}  [Y_d Y_u^\dagger   ]_{sr}+ N_c C^*_{\substack{H ud \\ rs}}  [Y_u Y_d^\dagger   ]_{rs} -2  C_{\substack{H l \\ rs}}^{(3)} [Y_e^\dagger  Y_e  ] _{sr}. \nonumber
\eea
The net result of the $\bar h$ dependence of $\Lambda$ (\ref{lamdep}) results in $\bar h^2/\Lambda^2$ having the profile of a `kick' that attains its maximum just before the Higgs decouples from all other SM fields. One might then imagine that processes that resulted in the terms in (\ref{bdef}) and (\ref{cdef}) might start to drop out of the running as $\bar h \to M_p/\xi$. Following Ref. \cite{DeSimone:2008ei}, one can roughly model this behaviour by multiplying each term containing an internal Higgs propagator by a factor of $s(\bar h)$, thus multiplying the $B_i$ by $s$ and the $D_i$ by $s^2$ in the above. (The factor $s(\bar h)$ should only really be applied to the singlet Higgs field.) The net effect of doing this, compared to simply scanning over the $A_i$ (i.e. ignoring the effects of terms with internal Higgs lines altogether) turns out to be negligible once we've scanned over the Wilson coefficients. 
This shouldn't be too surprising over the short range over which the RG effect we include has any support, the net effect of the $B_i$ and the $D_i$ can evidently simply be absorbed in to the $A_i$ Wilson coefficients. 

\bibliographystyle{JHEP}
\bibliography{RGCliff0506}

\end{document}